\documentclass[aps,showpacs,preprintnumbers,superscriptaddress,nofootinbib]{revtex4}
\usepackage{subfigure,graphicx,epsfig,amsmath,amsfonts,amssymb,xcolor,slashed,ulem}

\usepackage{float}
\usepackage{wrapfig}

\newcommand{\im}{{\rm Im}}
\newcommand{\re}{{\rm Re}}

\newcommand{\minv}{M_{\rm inv}}
\newcommand{\sumpol}{\sum_{\rm pol}}
\renewcommand\sout{\bgroup \color[rgb]{1,0,0} \ULdepth=-.5ex \ULset}
\allowdisplaybreaks[1]
\begin{document}
\title{Triangle singularities in $B^-\rightarrow K^-\pi^-D_{s0}^+$ and
$B^-\rightarrow K^-\pi^-D_{s1}^+$}
\date{\today}

\author{S.~Sakai}
\email{shuntaro.sakai@ific.uv.es}
\affiliation{Departamento de
F\'{\i}sica Te\'orica and IFIC, Centro Mixto Universidad de Valencia-CSIC Institutos de Investigaci\'on de Paterna, Aptdo.22085, 46071 Valencia, Spain}

\author{E.~Oset}
\affiliation{Departamento de
F\'{\i}sica Te\'orica and IFIC, Centro Mixto Universidad de Valencia-CSIC Institutos de Investigaci\'on de Paterna, Aptdo.22085, 46071 Valencia, Spain}

\author{A. Ramos}
\affiliation{Departament de F\'\i sica Qu\`antica i Astrof\'\i sica and Institut de Ci\`encies del Cosmos, Universitat de Barcelona, Mart\'i i Franqu\`es 1, 08028 Barcelona, Spain}

\begin{abstract} 

We study the appearance of structures in the decay of the $B^-$ into $K^- \pi^- D_{s0}^+(2317)$ and $K^- \pi^- D_{s1}^+(2460)$ final states by forming invariant mass distributions of $\pi^- D_{s0}^+$ and $\pi^- D_{s1}^+$ pairs, respectively. The structure in the distribution is associated to the kinematical triangle singularity that appears when the $B^- \to K^- K^{*\,0} D^0$ ($B^- \to K^- K^{*\,0} D^{*\,0}$) decay process is followed by the decay of the $K^{*\,0}$ into $\pi^- K^+$ and the subsequent rescattering of the $K^+ D^0$ ($K^+ D^{*\,0}$) pair forming the $D_{s0}^+(2317)$ ($D_{s1}^+(2460)$) resonance. We find this type of non-resonant peaks at 2850 MeV in the invariant mass of $\pi^- D_{s0}$ pairs from $B^- \to K^- \pi^- D_{s0}^+(2317)$ decays and around 3000 MeV in the invariant mass of $\pi^- D_{s1}^+$ pairs from $B^- \to K^- \pi^- D_{s1}^+(2460)$ decays. By employing the measured branching ratios of the
$B^- \to K^- K^{*\,0} D^0$ and $B^- \to K^- K^{*\,0} D^{*\,0}$ decays, we predict
the branching ratios for the processes $B^-$ into $K^- \pi^-
D_{s0}^+(2317)$ and $K^- \pi^- D_{s1}^+(2460)$, in the vicinity of the
triangle singularity peak, to be about $8\times10^{-6}$ and $1\times
10^{-6}$, respectively.
The observation of this reaction would also give extra support to the
molecular picture of the $D_{s0}^+(2317)$ and $D_{s1}^+(2460)$.

\end{abstract}


\maketitle

\section{Introduction}

Understanding the observed spectrum of hadron resonances \cite{pdg} and establishing a connection with the underlying theory of the strong interaction, Quantum Chromodynamics (QCD), is one of the prime 
goals of the hadron physics community.  In conventional quark models,
baryons are composed by three quarks, while mesons contain a quark and
an antiquark. However, even if these models provide a rather successful
description of a wealth of data for meson and baryon resonances
\cite{Godfrey:1985xj,Capstick:1986bm,Vijande:2004he}, other more exotic
components cannot be ruled out, especially considering that the degrees
of freedom of the QCD lagrangian contain not only quarks but also
gluons, which interact among each other as a consequence of the
non-abelian character of QCD. For many years, the intriguing possibility
of finding evidence for these exotic components in the mesonic and
baryonic spectrum has motivated a large amount of theoretical and
experimental activity, aiming at obtaining a quantitative understanding
of the confinement of quarks and gluons in QCD (for recent reviews on
this subject, see
Refs.~\cite{Klempt:2007cp,Crede:2008vw,Brambilla:2014jmp,Chen:2016qju,Guo:2017jvc}).

The advent of copious observation of hadron resonances, emerging as
peaks in the invariant mass distribution of selected hadrons from  the
decay of a heavy particle, requires a careful analysis of data, as some
of the resonance-like structures could be associated to, or affected by,
a triangle singularity of a Feynman diagram contributing to the
process. Early introduced by Landau \cite{Landau:1959fi}, this kinematic
singularity occurs if the diagram involves three intermediate-state
particles which can be placed simultaneously on-shell with their momenta
being collinear in the frame of reference of the decaying particle.
These especial conditions permit to fuse two of the loop particles into
an external outgoing one, hence allowing for an interpretation of the
Feynman diagram as a classical process, which is the essence of the
Coleman-Norton theorem \cite{Coleman:1965xm} and corresponds to placing
the singularity on the physical boundary producing an observable effect.
It is worth pointing out that a triangle singularity is favored when
the two loop hadrons fuse into a hadron molecule mainly composed of
these hadrons, as molecules are generally formed close to
threshold, hence favoring the fulfillment of the on-shell conditions.

A clear manifestation of a triangle singularity is found in the decays
of the $\eta(1405)$. It has been shown that the unusually large ratio of
rates between the isospin-violating decay,  $\eta(1405) \to \pi^0
f_0(980) \to \pi^0,\pi^+ \pi^-$, and the
process $\eta(1405) \to \pi^0 a_0(980) \to \pi^0,\pi^0 \eta$ found by BESIII \cite{BESIII:2012aa} can be naturally explained in terms of a triangle singularity \cite{Wu:2011yx}. The isospin breaking rate is enhanced by the sequence of  processes involved in the triangle diagram, namely  the $\eta(1405)$ decaying into a $K^* {\bar K}$ pair, followed by the decay of the $K^*$ into $K \pi$ and the subsequent fusion of the $K$ with the ${\bar K}$ to form the $f_0(980)$.  This explanation was corroborated in Ref.~\cite{Aceti:2012dj}, where the absolute value for the ratio $\Gamma(\eta(1405)\to  \pi^0,\pi^+ \pi^-)/\Gamma(\eta(1405)\to  \pi^0,\pi^0 \eta)$ was obtained, and it was found, together with the line shapes of the two reactions, to be in good agreement with experiment.  One finds a similar example in the signals reported by COMPASS, from the scattering of high energy pions off protons  \cite{Adolph:2015pws}. A peak in the distribution, which could be easily interpreted as the sign of an axial-vector $a_1(1420)$,  can naturally receive an explanation in terms of a triangle singularity  involving a virtual  $a_1(1260)$ which decays into $K^* {\bar K}$, followed by the decay of the $K^*$ into $K\pi$, where the $\pi$ is emitted and the $K$ merges with the ${\bar K}$ forming the $f_0(980)$. This was noted by the authors of  Ref.~ \cite{Liu:2015taa} and explicitly evaluated in \cite{Ketzer:2015tqa},  finding a good description of the experimental facts observed in  \cite{Adolph:2015pws}. A confirmation of this mechanism was offered by the work of Ref.~\cite{Aceti:2016yeb}, where the interference between the $K^* K{\bar K}$ and $\rho\pi\pi$ loops was also analyzed.

In the heavy flavor sector, some of the claimed exotic hadrons are also being analyzed on the basis of triangle singularity effects. The possibility that the narrow peak associated to the $P_c(4450)$ pentaquark by the LHCb collaboration \cite{Aaij:2015tga} could be associated to a triangle singularity was explored in 
\cite{Guo:2015umn,Liu:2015fea,Guo:2016bkl}. A thorough study of possible charmonium-$\Lambda^*$ pairs contributing to this signal was carried out in Ref.~ \cite{Bayar:2016ftu}, finding the $\chi_{c1}\Lambda(1890)$ pair as the best candidate if the pentaquark had $J^P=1/2^+$ or $3/2^+$, but not being able to reproduce the experimental features if the quantum numbers are $J^P=3/2^-$ or $5/2^+$, as preferred by the experiment for the narrow peak observed. The existence of some exotic meson states, such as the charged charmonium $Z_c(3900)$ \cite{Ablikim:2013mio,Ablikim:2013emm,Liu:2013dau,Xiao:2013iha}, has also been challenged in favor of a triangle singularity explanation \cite{Wang:2013cya,Liu:2013vfa,Liu:2015taa}, as well as for other quarkonium \cite{Liu:2014spa} and bottomonium \cite{Wang:2013hga} states. However,  in general, no conclusion can be drawn on whether the signal should be associated to a triangular singularity rather than to a resonance-pole, or even on quantifying possible interferences among these two mechanisms, unless the strength of the triangle diagram can be determined in terms of experimentally known vertices. 
Attempts to parametrize amplitudes in terms of triangular singularity and
other elements, as done in Ref.~\cite{pilloni}, and conducting fits to high
quality data, could help in the future.

It is also useful to predict the location of triangular singularity signals in processes that may easily be produced in present experimental facilities, in order to anticipate their nature. A recent example is the decay $B_c \to B_s \pi \pi$, which has been investigated via the ${\bar K}^*B{\bar K}$ loop in Ref.~\cite{Liu:2017vsf} and has produced a $X(5777)$ structure in the invariant mass distribution of  $B_s^0 \pi^+$ states.  This process perfectly fulfills the triangle singularity condition and it has the advantage that the interaction of the $B{\bar K}$ in the loop, transforming into the final $B_s^0 \pi^+$, is too weak to produce a resonance, hence the structure observed would be of pure kinematical origin. 

In the present work, we also study the appearance of pure kinematic
structures in the decay of the $B^-$ into $K^- \pi^- D_{s0}^+(2317)$ and
$K^- \pi^- D_{s1}^+(2460)$ final states by forming invariant mass
distributions of $\pi^- D_{s0}^+$ and $\pi^- D_{s1}^+$ pairs, respectively.
{The $D_{s0}^+(2317)$ was first observed in the BaBar collaboration
as a narrow peak in the $D_s\pi$ invariant mass distribution \cite{new1}.
The state was confirmed by CLEO \cite{new2} and Belle \cite{new3}.
Nowadays is already well established in the PDG \cite{pdg} with quantum
numbers $I(J^P)=0(0^+)$.
The $D_{s1}^+(2460)$ was also observed in the CLEO experiment
\cite{new2} in the $D_s^*\pi$ channel and BaBar also found a signal in
that region.
Nowadays it is also well established in the PDG \cite{pdg} with quantum
numbers $0(1^+)$.
On the theoretical side, there has been much work claiming that these
states are of molecular nature, mostly $DK$ for the $D_{s0}^+(2317)$ and
$D^*K$ for the $D_{s1}^+(2460)$
\cite{Barnes:2003dj,vanBeveren:2003kd,kolo,hofmann,chiang,newguo,Gamermann:2006nm}.
Support for this picture in several reactions has also been discussed in
Refs.~\cite{Faessler:2007gv,Guo:2008gp,Guo:2009ct,Cleven:2014oka}.
Lattice QCD calculations have also
{found indications for this interpretation}
\cite{lattice1,lattice2}, and a reanalysis of these latter results
\cite{sasa} even provides  the amount of $DK$ and $D^*K$ component in the wave function of
the $D_{s0}^+(2317)$ and $D_{s1}^+(2460)$, respectively.
The $B^-\rightarrow K^-\pi^-D_{s0}^+(2317)$ and $B^-\rightarrow
K^-\pi^-D_{s1}^+(2460)$}
processes occur via 
$K^* \, K\, D$ or $K^* \, K\, D^*$ intermediate loops, which meet the condition of the triangle singularity when the $K D$ or $KD^*$ fuse into the $D_{s0}^+(2317)$ or $D_{s1}^+(2460)$, respectively. We find a peak at 2850 MeV in the invariant mass of $\pi^- D_{s0}^+$ pairs from the $B^- \to K^- \pi^- D_{s0}^+(2317)$ decay and around 3000 MeV in the invariant mass of $\pi^- D_{s1}^+$ pairs from the $B^- \to K^- \pi^- D_{s1}^+(2460)$ process. We also quantify the strength under these peaks  by explicitly evaluating the triangle singularity diagrams corresponding to these processes and referring them to the decays  $B^-\rightarrow K^-K^{*0}D^0$ and $B^-\rightarrow K^-K^{*0}D^{*0}$, whose branching ratios are reported in the PDG \cite{pdg}. We
predict
the branching ratios for the processes $B^-$ into $K^- \pi^-
D_{s0}^+(2317)$ and $K^- \pi^- D_{s1}^+(2460)$ in the vicinity of the
triangle singularity peak to be about $8\times10^{-6}$ and $1\times
10^{-6}$, respectively.

\section{Formalism}

\begin{figure}[ht]
 \centering
 \includegraphics[width=8cm]{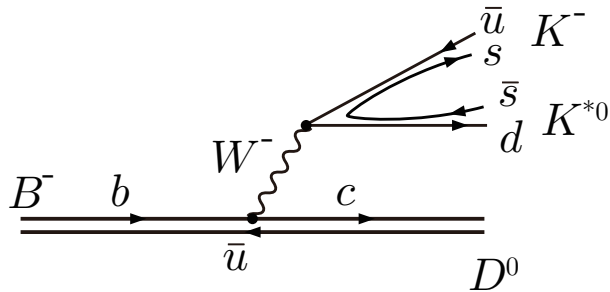}
 \caption{Quark picture for the $B^-\rightarrow K^-K^{\ast0}D^0$ decay.}
 \label{fig_1}
\end{figure}

Let us first start with  the $B^-\rightarrow
K^-K^{\ast0}D^0$ decay process, which can be visualized in a quark representation in
Fig.~\ref{fig_1}.
The process is Cabibbo favored with the topology of external emission
\cite{chau,chaucheng} and, hence, also favored by the color counting.
In the $W^-$ external emission, a $\bar{u}d$ pair is formed, which is
hadronized with a $\bar{q}q$ pair with vacuum quantum numbers, and
through the $\bar{s}s$ component, gives rise to $K^-$ and a $K^{\ast0}$.
The $c\bar{u}$ final state in Fig.~\ref{fig_1} gives rise to the $D^0$,
thus completing the final state.
The branching ratio for the $B^-\rightarrow K^-K^{*0}D^0$ decay is given in the
PDG~\cite{pdg} as a BR$(B^-\rightarrow K^-K^{*0}D^0)=(7.5\pm 1.7)\times
10^{-4}$ measured by Belle in Ref.~\cite{belle}.
Similarly, we can also have, with the same topology as in
Fig.~\ref{fig_1}, the $B^-\rightarrow K^-K^{*0}D^{*0}$ decay, which is
reported in the PDG with a branching ratio of BR$(B^-\rightarrow
K^-K^{*0}D^{*0})=(1.5\pm 0.4)\times10^{-3}$, also measured in
Ref.~\cite{belle}.

\subsection{$B^-\rightarrow K^-\pi^-D_{s0}^+(2317)$ decay}
\label{sec_2_1}

\begin{figure}[th]
 \centering
 \includegraphics[width=8cm]{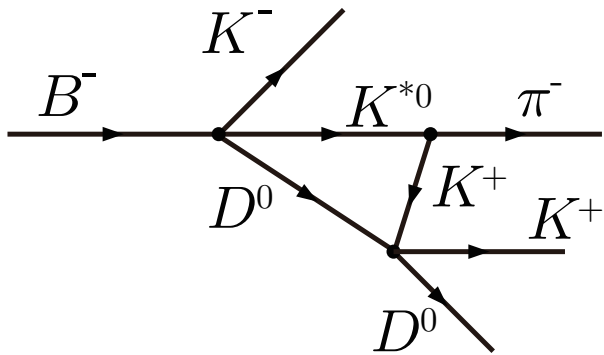}
 \caption{Triangle diagram stemming from $B^-\rightarrow K^-K^{*0}D^0$,
 with $K^{*0}\rightarrow\pi^-K^+$ and $D^0K^+$ rescattering.}
 \label{fig_2}
\end{figure}

\begin{figure}[th]
 \centering
 \includegraphics[width=8cm]{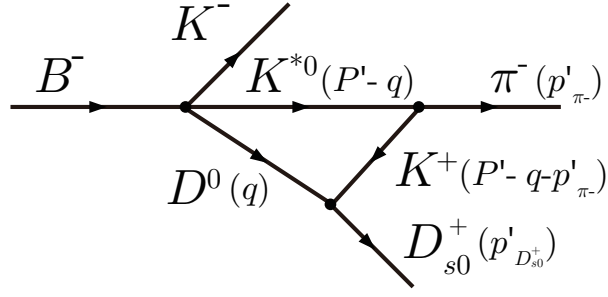}
 \caption{Triangle diagram for the $B^-\rightarrow K^-\pi^-D_{s0}^+$
 process.
 In parenthesis the momenta of the particles.}
 \label{fig_3}
\end{figure}

From the process $B^-\rightarrow K^-K^{*0}D^0$ one can form the triangle
mechanism depicted in Fig.~\ref{fig_2}, which was found in
Ref.~\cite{Liu:2015taa} to develop a singularity.
However, the energy at which the singularity appears depends on the
invariant mass of the $D^0K^+$ pair, and is therefore diluted over the phase space of
$D^0K^+$.
Hence, in order to show a neat singular peak, it is preferable to
concentrate the $D^0K^+$ pair strength in one resonance. For this
purpose, the $D_{s0}^+(2317)$ is the obvious choice, since
this resonance is to a good approximation a $DK$ molecule in $I=0$
\cite{kolo,hofmann,chiang,Gamermann:2006nm,sasa} and hence it couples strongly
to $DK$ states.
Thus, we choose to study the process represented in Fig.~\ref{fig_3}.

In other cases of triangle singularities \cite{Liu:2015taa,Guo:2015umn},
one starts with the decay of a particle into three final ones, and then one
varies the invariant mass of two final particles, as e.g. those emitted in the lower vertex of the triangle, to see for which invariant mass of this pair the singularity appears. However, if this strategy is applied with four particles in the final
state, as in Fig.~\ref{fig_2}, the effect of the triangle singularity
would be concentrated in a small region of the whole phase space.
Therefore, by producing a resonance from the $DK$ pair, one is effectively rendering the
problem into one of three particles in the final state, with a smaller
phase space, for which the singularity has a larger weight.
 
The evaluation of the diagram in Fig.~\ref{fig_3} requires first to
provide an expression for the $B^-\rightarrow K^-K^{*0}D^0$
vertex. Because a $P$-wave coupling is needed in this process to
conserve angular momentum, we can write the corresponding amplitude as follows:
\begin{align}
 -{\rm i}\,t_{B^-,K^-K^{*0}D^0}=-{\rm i}\,C\,\vec{\epsilon}_{K^{*0}}\,\vec{{p}}_{K^-}\label{eq_gbmkmks0d0} \ ,
\end{align}
where $\vec{{p}}_{K^-}$ is the $K^{-}$ momentum in a suitable frame and $\vec{\epsilon}_{K^{*0}}$ is the polarization vector of the $K^{*0}$.
The choice of this vertex, involving the $\vec{p}_{K^-}$ momentum, requires a justification.
Let us note that the two vertices involved in the exchange of the $W^-$ in Fig.~\ref{fig_1} lead to an operator of the type
$\gamma^\mu(1-\gamma_5) \gamma_\mu(1-\gamma_5)$ 
 at the quark level, neglecting the term inversely proportional to the large mass squared of the $W^-$ propagator. The big contributions of this term will be associated to the operators $\gamma^0$ and $\gamma^i\gamma_5$,
which in the nonrelativistic limit correspond to $1$ and $\sigma^i$, respectively. The term proportional to $\sigma^i \sigma^i$ will not contribute because, by virtue of the  Wigner-Eckart theorem, the expectation value of the $\sigma^i$ operator between the two $0^-$ states, corresponding to $B^-$ and $D^0$ at the lower vertex,
vanishes. Moreover, we recall that the
 $\bar{q}q$ pair at the other vertex, involved in the hadronization process, has spin $S=1$
and $L=1$, according to the $^3P_0$ model \cite{micu,yaouanc}.
The $S=1$ carried by the  $\bar{q}q$ pair at the quark level allows one to produce the vector meson and, therefore, an operator of the
type $\epsilon_{K^*\mu}(p_K-p_{K^*})^\mu$ is expected at the hadronic level in the rest frame of the $K^-K^{*0}$ pair, where $(p_K-p_{K^*})^\mu$ stands for the relative four-momentum, as required by the $L=1$ of the $\bar{q}q$ pair initiating the hadronization process.
The Lorenz condition $\epsilon_{K^*\mu}p^\mu_{K^*}=0$ allows one  to finally write the vertex as 
$\epsilon_{K^*\mu}p^\mu_{K}$, which is the covariant form of
Eq.~(\ref{eq_gbmkmks0d0}) that will also be used later on.

The position of the triangle singularity is of kinematical origin and should not depend strongly on the specific details of the weak decay vertex employed. However, in sect.~\ref{sec:reson} we will also explore how the results are modified if the decay vertex assumes the process $B^-\rightarrow K^-K^{*0}D^{(*)\,0}$ to be dominated by the subthreshold $a_1(1260)$ resonance coupling strongly to the $K^-K^{*0}$ pair. This will modify the strength of the singularity, as we will see.

We shall see that a singularity appears around 2815~MeV for the
$\pi^-D_{s0}^+(2317)$ invariant mass in the $B^-\rightarrow K^-\pi^-D_{s0}^+$ process.
This means that, in the $\pi^-D_{s0}^+(2317)$ rest frame where we will
evaluate the triangle diagram, the $K^{*0}$ momentum is about
262~MeV$/c$, which is small compared to the $K^{\ast 0}$ mass. This allows us to neglect the time component $\epsilon^0$ of the $K^{\ast 0}$ polarization vector,
and employ the following formula for the polarization sum,
\begin{align}
 \sumpol{\epsilon}_{K^{*0}i}{\epsilon}_{K^{*0}j}=&\delta_{ij} \ , \label{eq_sum_pol} 
\end{align}
when required. The parameter $C$ in Eq.~(\ref{eq_gbmkmks0d0}) can be
related to the partial decay rate of the $B^-\rightarrow K^-K^{*0}D^0$
process, which is given by
\begin{align}
 \Gamma_{B^-\rightarrow K^-K^{*0}D^0}=\int
 d\minv(D^0K^{*0})&\frac{1}{(2\pi)^3}\frac{|\vec{\tilde{p}}_{K^-}||\vec{\tilde{p}}^\prime_{K^{*0}}|}{4m_{B^-}^2}\sumpol\left|t_{B^-,K^-K^{*0}D^0}\right|^2,
 \label{eq:gamma1}
\end{align}
where $\minv(D^0K^{*0})$ is the $D^0K^{*0}$ invariant mass,
$\vec{\tilde{p}}_{K^-}$ and $\vec{\tilde{p}}^\prime_{K^{*0}}$ are the momenta of the
$K^-$ in the $B^-$ rest frame and that of the $K^{*0}$ in the $D^0K^{*0}$
center-of-mass (CM) frame, respectively, given by
\begin{align}
 |\vec{\tilde{p}}_{K^-}|=&\frac{1}{2m_{B^-}}\lambda^{1/2}(m_{B^-}^2,m_{K^-}^2,\minv^2(D^0K^{*0})),\\
 |\vec{\tilde{p}}\,^\prime_{K^{*0}}|=&\frac{1}{2\minv(D^0K^{*0})}\lambda^{1/2}(\minv^2(D^0K^{*0}),m_{K^{*0}}^2,m_{D^0}^2), 
\end{align}
where $\lambda(x,y,z)=x^2+y^2+z^2-2xy-2yz-2zx$ is the K\"allen
function. From the vertex of Eq.~(\ref{eq_gbmkmks0d0}), which assumes
the  $\epsilon^0$ component of the $K^*$ polarization
{to give a small contribution}, and employing the polarization sum of Eq.~(\ref{eq_sum_pol}), we have
\begin{align}
 \sumpol | t_{B^-,K^-K^{*0}D^0}|^2=C^2\,|\vec{\tilde{p}}\,^\prime_{K^-}|^2\label{const_t_1}
\end{align}
written in terms of the $K^-$ momentum in the
$D^0K^{*0}$ CM frame.
However, the phase space for $B^-\rightarrow K^-D^0K^{*0}$ decay is such
that the invariant mass of $D^0K^{*0}$ peaks at 3100~MeV, where
$|\vec{\tilde{p}}\,'_{K^{*0}}|=670$ MeV$/c$ and stretches at even higher invariant masses.
Therefore, $|\vec{\tilde{p}}\,'_{K^{*0}}|/m_{K^{*0}}$ is not small, and one must employ
the covariant form of
Eq.~(\ref{eq_gbmkmks0d0}),
\begin{align}
 -{\rm i}t_{B^-,K^-K^{*0}D^0}=&{\rm i}C\epsilon_{K^{*0}\mu}p^\mu_{K^-} \ .\label{eq_tbmdec_rel}
\end{align}
The sum over polarizations in Eq.~(\ref{const_t_1}), evaluated in the $D^0K^{*0}$ rest frame and
after performing an angular integration over the $K^-$ and $K^{*0}$ angle in that
frame, becomes then
\begin{align}
 \sumpol|t_{B^-,K^-K^{*0}D^0}|^2=C^2\left(-m_{K^-}^2+\frac{(\tilde{p}'^0_{K^-}\tilde{p}'^0_{K^{*0}})^2+\frac{1}{3}(|\vec{\tilde{p}}\,'_{K^-}||\vec{\tilde{p}}\,'_{K^{*0}}|)^2}{m_{K^{*0}}^2}\right),
  \label{eq:rel_vertex}
\end{align}
 where $|\vec{\tilde{p}}\,^\prime_{K^-}|$ is given by
\begin{align}
 |\vec{\tilde{p}}\,^\prime_{K^-}|=\frac{1}{2\minv(D^0K^{*0})}\lambda^{1/2}(m_{B^-}^2,m_{K^-}^2,\minv^2(D^0K^{*0})) \ ,
\end{align}
and $\tilde{p}^{\prime0}_{K^-}$ and $\tilde{p}^{\prime0}_{K^{*0}}$ are given by
$\sqrt{m_{K^-}^2+|\vec{\tilde{p}}\,^\prime_{K^-}|^2}$ and
$\sqrt{m_{K^{*0}}^2+|\vec{\tilde{p}}\,^\prime_{K^{*0}}|^2}$, respectively.
Thus,
{inserting Eq.~(\ref{eq:rel_vertex}) into Eq.~(\ref{eq:gamma1}), one
obtains $C^2$ as}
\begin{align}
 C^2=\frac{\Gamma_{B^-\rightarrow K^-K^{*0}D^0}}{\displaystyle\int
 d\minv(D^0K^{*0})\displaystyle\frac{1}{(2\pi)^3}\displaystyle\frac{|\vec{\tilde{p}}_{K^-}||\vec{\tilde{p}}\,'_{K^{*0}}|}{4m_{B^-}^2}\left(-m_{K^-}^2+\frac{(\tilde{p}'^0_{K^-}\tilde{p}'^0_{K^{*0}})^2+\frac{1}{3}(|\vec{\tilde{p}}\,'_{K^-}||\vec{\tilde{p}}\,'_{K^{*0}}|)^2}{m_{K^{*0}}^2}\right)}.\label{eq_csq}
\end{align}

Next, we look into the $K^{*0}\pi^- K^+$ vertex in Fig~\ref{fig_3}.
We can obtain this $VPP$ vertex from the chiral invariant lagrangian with
local hidden symmetry given in Refs.~\cite{hidden1,hidden2,hidden4,Nagahiro:2008cv},
\begin{align}
 \mathcal{L}_{VPP}=&-{\rm i}\,g\left<V^\mu[P,\partial_\mu P]\right> \ ,
\end{align}
where the brackets $\left<...\right>$ stand for the trace of
the flavor SU(3) matrices and $g$ is the coupling in the local hidden
gauge, $g=m_V/2f_\pi$, with $m_V=800$~MeV and $f_\pi=93$~MeV.
The symbols $V_\mu$ and $P$ stand for the octet vector meson field and the octet pseudoscalar meson
field SU(3) matrices, respectively (see, $e.g.,$ Ref.~\cite{Nagahiro:2008cv} for their explicit forms).
The amplitude of the $K^{*0}$ decay into $K^+\pi^-$ is then
\begin{align}
 -{\rm i}\,t_{K^{*0},K^+\pi^-}=&-{\rm i}\,g\, \vec{\epsilon}_{K^{*0}}(\vec{p}\,^{\prime}_{\pi^-}-\vec{p}\,^{\prime}_{K^+}) \ ,\label{eq_gkskpi}
\end{align}
where $\vec{p}\,^{\prime}_{\pi^-}$ and $\vec{p}\,^{\prime}_{K^+}$ are the momenta of the
$\pi^-$ and ${K^+}$, respectively, in the chosen reference frame of
$\pi D_{s0}^{+}$ at rest, where $\epsilon^0_{K^{*0}}$ can be neglected. 

Finally, for the evaluation of the triangle diagram in Fig.~\ref{fig_3},
we also need to provide the $D_{s0}^+ D^0 K^+$ vertex. We follow the
model of Ref.~\cite{Gamermann:2006nm}, where the $D_{s0}^+$ is found  to be dynamically
generated from the $DK$, $D_s\eta$, and $D_s\eta_c$ interaction in $S$-wave within a chiral unitary approach. The $D_{s0}^+ D^0K^+$ vertex is then given by
\begin{align}
 -{\rm i}\,t_{D_{s0}^+,D^0K^+}=-{\rm i}\,g_{D^+_{s0},D^0K^+} \ ,\label{eq_gds0d0kp}
\end{align}
where the coupling of the $D^+_{s0}$ to  $D^0K^+$ states,
$g_{D_{s0}^+,D^0K^+}$, is obtained from the coupling constant of the $D_{s0}$ to the $DK$ channel in isospin $I=0$,
found to be $g_{D_{s0},DK}=10.21$ GeV in
Ref.~\cite{Gamermann:2006nm}, multiplied by the appropriate Clebsch-Gordan (CG) coefficient, namely
$g_{D_{s0}^+,D^0K^+}=-g_{D_{s0},DK}/\sqrt{2}$.

With the vertices of Eqs.~(\ref{eq_gbmkmks0d0}), (\ref{eq_gkskpi}),
and (\ref{eq_gds0d0kp}), the amplitude of the triangle diagram of
Fig.~\ref{fig_3}, evaluated in the $\pi^-D_{s0}^+$ CM frame, is given by
\begin{align}
 t_{B^-,K^-\pi^-D_{s0}^+}=&\,C\,g\,g_{D_{s0}^+,D^0K^+}\sumpol
{\rm i}\int\frac{d^4q}{(2\pi)^4} \, \left(\vec{\epsilon}_{K^{*0}}\,\vec{p}\,^\prime_{K^-}\right) \, \left(\vec{\epsilon}_{K^{*0}}\, (2\vec{p}\,^\prime_{\pi^-}+\vec{q}\,)\right)\notag\\
 &\frac{1}{q^2-m_{D^0}^2+{\rm i}\epsilon}\frac{1}{(P^\prime-q)^2-m_{K^{*0}}^2+{\rm i}\epsilon}\frac{1}{(P^\prime-q-p_{\pi^-}^\prime)^2-m_{K^+}^2+{\rm i}\epsilon}\notag \\
 \equiv& \, C\,g\,g_{D_{s0}^+,D^0K^+}\, t_T,\label{eq_t_1}
\end{align}
where the last identity defines $t_T$. In the former equation,
$P^\prime\equiv p^\prime_{\pi^-}+p^\prime_{D_{s0}^+}$, 
$p^\prime_{\pi^-}$ and $p^\prime_{K^-}$ are the momenta in the CM frame of the $\pi^-D_{s0}^+$ pair. Hence,
$P^{\prime 0}=\minv(\pi^-D^+_{s0})$, $\vec{P}\,^\prime=\vec{0}$, 
\begin{align}
|\vec{p}\,^\prime_{\pi^-}|=\frac{1}{2\minv(\pi^-D^+_{s0})}\lambda^{1/2}(\minv^2(\pi^-D^+_{s0}),m_{\pi^-}^2,m_{D_{s0}^+}^2) \ ,
\label{eq:pimom}
\end{align}
and
\begin{align}
|\vec{p}\,^\prime_{K^-}|=\frac{1}{2\minv(\pi^-D^+_{s0})}\lambda^{1/2}(m_{B^-}^2,m_{K^-}^2,\minv(\pi^-D^+_{s0})^2) \ .
\label{eq:kamom}
\end{align}

Using Eq.~(\ref{eq_sum_pol}), the sum over the polarizations  of the $K^{*0}$  gives
\begin{align}
 \sumpol \left(\vec{\epsilon}_{K^{*0}}\,\vec{p}\,^\prime_{K^-}\right) \, \left(\vec{\epsilon}_{K^{*0}} \, (2\vec{p}\,^\prime_{\pi^-}+\vec{q}\,)\right)=\vec{p}\,^\prime_{K^-}(2\vec{p}\,^\prime_{\pi^-}+\vec{q}\,)\ ,
\end{align}
and, then, the function $t_T$ defined in Eq.~(\ref{eq_t_1}) can be written as
\begin{align}
 t_T=&{\rm i}\int\frac{d^4q}{(2\pi)^4}\,\vec{p}\,^\prime_{K^-}(2\vec{p}\,^\prime_{\pi^-}+\vec{q}\,)
 \frac{1}{q^2-m_{D^0}^2+{\rm i}\epsilon}
 \frac{1}{(P^\prime-q)^2-m_{K^{*0}}^2+{\rm i}\epsilon}\frac{1}{(P^\prime-q-p^\prime_{\pi^-})^2-m_{K^+}^2+{\rm i}\epsilon} \ .
 \label{eq_tT3}
\end{align}
We can perform the $q^0$ integration analytically, in the same way as shown in
Refs.~\cite{Aceti:2015zva,Bayar:2016ftu,Wang:2016dtb}.
As for the three-momentum integral we note that, since the only three-momentum that is not integrated in Eq.~(\ref{eq_tT3}) is
$\vec{p}\,^\prime_{\pi^-}$, one may employ the following identity 
\begin{align}
\int d^3q\, q_i \, f(\vec{q}\,,\vec{k}^\prime)=k^\prime_i \int
d^3q\frac{ (\vec{q}\, \vec{k}^\prime)}{|\vec{k}^\prime|^2} f(\vec{q}\,,\vec{k}^\prime) \ ,
\end{align}
with $\vec{k}^\prime=\vec{p}\,^\prime_{\pi^-}$ and $f(\vec{q}\,,\vec{k}^\prime)$ being
the product of the $D^0$, $K^{*0}$, and $K^+$
propagators. Hence, we obtain:
\begin{align}
 t_T=&(\vec{p}\,^\prime_{K^-} \vec{p}\,^\prime_{\pi^-})\int\frac{d^3q}{(2\pi)^3}\frac{1}{8\omega(\vec{q}\,)\omega^\prime(\vec{q}\,)\omega^*(\vec{q}\,)}
 \,\frac{1}{p^{\prime0}_{\pi^-} -\omega^\prime(\vec{q}\,) -\omega^*(\vec{q}\,) +{\rm i}\epsilon}
 \,\frac{1}{P^{\prime0}-\omega^*(\vec{q}\,)-\omega(\vec{q}\,)+{\rm i}\epsilon}\notag\\
 &\,\frac{2P^{\prime0}\omega(\vec{q}\,) +2p^{\prime0}_{\pi^-}\omega^\prime(\vec{q}\,)
 -2[\omega(\vec{q}\,)+\omega^\prime(\vec{q}\,)][\omega(\vec{q}\,) +\omega^\prime(\vec{q}\,) +\omega^*(\vec{q}\,)]}{(P^{\prime0}
 -\omega(\vec{q}\,) -\omega^\prime(\vec{q}\,) -p^{\prime0}_{\pi^-} +{\rm i}\epsilon)(P^{\prime0} +\omega(\vec{q}\,) +\omega^\prime(\vec{q}\,)
 -p^{\prime0}_{\pi^-}
 -i\epsilon)}\left(2+\frac{\vec{p}\,^\prime_{\pi^-} \vec{q}\,}{|\vec{p}\,^\prime_{\pi^-}|^2}\right)\notag\\
 \equiv&(\vec{p}\,^\prime_{K^-} \vec{p}\,^\prime_{\pi^-})\, \tilde{t}_T,\label{eq_tT1}
\end{align}
where the last identity defines the triangle singularity integral $\tilde{t}_T$ out of the amplitude $t_T$. We recall that, in the integrand of Eq.~(\ref{eq_tT1}),
$P^{\prime0}$ denotes the invariant mass of the $\pi^-D_{s0}^+$ pair,
$\minv(\pi^-D^+_{s0})$, while
 $\omega(\vec{q}\,)=\sqrt{m_{D^0}^2+|\vec{q}\,|^2}$ is the energy of the $D^0$,
$\omega^*(\vec{q}\,)=\sqrt{m_{K^{*0}}^2+|\vec{q}\,|^2}$ that of the $K^{*0}$,
 $\omega^\prime(\vec{q}\,)=\sqrt{m_{K^+}^2+|\vec{q}\,+\vec{p}\,^\prime_{\pi^-}|^2}$  that of the $K^+$,
and
$p^{\prime0}_{\pi^-}=\sqrt{m_{\pi^-}^2+|\vec{p}\,^\prime_{\pi^-}|^2}$ that of the $\pi^-$ in the $\pi^-D_{s0}^+$ CM frame.
The $|\vec{q}\,|$ integral in Eq.~(\ref{eq_tT1}) is regulated by a cutoff
$|\vec{q}\,|_{\rm max}=800$~MeV in the $D^+_{s0}$
rest frame, which reproduces the results obtained in the chiral unitary
approach for the $D_{s0}^+$ meson \cite{Gamermann:2006nm}.
We implement the width of the $K^{*0}$ with the replacement of
$\omega^\ast$ by $\omega^\ast-{\rm i}\Gamma_{K^{*0}}/2$ in the integrand of Eq.~(\ref{eq_tT1}).

Taking into account the amplitude of Eq.~(\ref{eq_t_1}), with $t_T$ given in Eq.~(\ref{eq_tT3}), and performing
the integration over the angle formed by the $K^-$ and
$\pi^-$ momenta, the mass distribution of  the $B^-\rightarrow
K^-\pi^-D_{s0}^+$ decay width is given by
\begin{align}
 \frac{d\Gamma_{B^-\rightarrow K^-\pi^-D_{s0}^+}}{d\minv(\pi^-
 D^+_{s0})}=&\frac{1}{(2\pi)^3}\frac{1}{4m_{B^-}^2}|\vec{p}_{K^-}||\vec{p}\,^\prime_{\pi^-}|\,\frac{C^2}{3}\,|\bar{t}^{\rm
 eff}_{B^-,K^-\pi^-D_{s0}^+}|^2,\label{eq_mass_dist_0}
\end{align}
where $\vec{p}_{K^-}$ is the momentum of the $K^-$ in the $B^-$ rest frame,
\begin{align}
 |\vec{p}_{K^-}|=&\frac{1}{2m_{B^-}}\lambda^{1/2}(m_{B^-}^2,m_{K^-}^2,\minv^2(\pi^-
 D_{s0}^+))\ ,
\end{align}
$\vec{p}\,^\prime_{\pi^-}$ is that of the $\pi^-$ in the CM frame of the $\pi^-D_{s0}^+$ pair [see Eq.~(\ref{eq:pimom})], and we have defined
$\bar{t}^{\rm
eff}_{B^-,K^-\pi^-B_{s0}}\equiv g\,g_{D_{s0}^+,D^0K^+}\,|\vec{p}\,^\prime_{K^-}||\vec{p}\,^\prime_{\pi^-}|\tilde{t}_T$.

Dividing both sides of Eq.~(\ref{eq_mass_dist_0}) by $\Gamma_B$ and using
Eq.~(\ref{eq_csq}), we obtain
\begin{align}
 \frac{1}{\Gamma_B}\frac{d\Gamma_{B^-\rightarrow
 K^-\pi^-D_{s0}^+}}{d\minv(\pi^-
 D_{s0}^+)}=&\frac{{\rm BR}(B^-\rightarrow
 K^-K^{*0}D^0)\,|\vec{p}_{K^-}||\vec{p}\,^\prime_{\pi^-}|\,\frac{1}{3}|\bar{t}^{\rm
 eff}_{B^-,K^-\pi^-D_{s0}^+}|^2}{\displaystyle\int
 d\minv(D^0K^{*0})|\vec{\tilde{p}}_{K^-}||\vec{\tilde{p}}'_{K^{*0}}|\left(-m_{K^-}^2+\frac{(\tilde{p}'^0_{K^-}\tilde{p}'^0_{K^{*0}})^2+\frac{1}{3}(|\vec{\tilde{p}}'_{K^-}||\vec{\tilde{p}}'_{K^
 {*0}}|)^2}{m_{K^{*0}}^2}\right)},\label{eq_mass_dist_1}
\end{align}
where the branching ratio ${\rm
BR}(B^-\rightarrow K^-K^{*0}D^0)=\Gamma_{B^-\rightarrow
K^-K^{*0}D^0}/\Gamma_B$ is $(7.5\pm 1.7)\times 10^{-4}$  \cite{pdg}.
Integrating Eq.~(\ref{eq_mass_dist_1}) over $\minv(\pi^-D^+_{s0})$, we
obtain the branching ratio for the $B^-\rightarrow K^-\pi^-D_{s0}^+$ process.

\subsection{$B^-\rightarrow K^-\pi^-D_{s1}^+(2460)$ decay}
\begin{figure}[hbt]
 \centering
 \includegraphics[width=8cm]{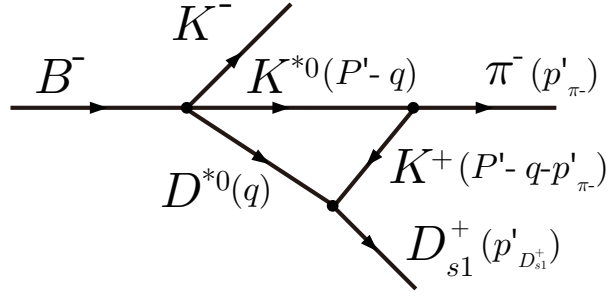}
 \caption{Triangle diagram for $B^-\rightarrow K^-\pi^-D_{s1}^+$.}
 \label{fig_4}
\end{figure}

The triangle diagram for the $B^-\rightarrow K^-\pi^-D_{s1}^+$ decay mode
is shown in Fig.~\ref{fig_4}.
We can evaluate the branching ratio of this process in a similar way
to that in Sec.~\ref{sec_2_1} for the $B^-\rightarrow K^-\pi^-D_{s0}^+$ process.
The differences from the previous case
are the amplitudes $B^-\rightarrow K^-K^{*0}D^{*0}$ and
$D^{*0}K^+\rightarrow D_{s1}^+$.

The first vertex is related to the decay $B^-\rightarrow K^-K^{*0}D^{*0}$, which can proceed in
$S$-wave. The
corresponding amplitude, $t_{B^-,K^-K^{*0}D^{*0}}$, in a suitable reference frame is then given by
\begin{align}
 -{\rm i}\,t_{B^-,K^-K^{*0}D^{*0}}=&-{\rm i}\,C^\prime\,\vec{\epsilon}_{K^{*0}}\,\vec{\epsilon}_{D^{*0}},\label{eq_gbmkmks0ds0}
\end{align}
where $\vec{\epsilon}_{D^{*0}}$  is the polarization vector  of the $D^{*0}$.
We fix the coupling constant $C^\prime$ in Eq.~(\ref{eq_gbmkmks0ds0}) from the branching ratio of the
$B^-\rightarrow K^-K^{*0}D^{*0}$ process.
The partial width for this decay is written as
\begin{align}
 \Gamma_{B^-\rightarrow K^-K^{*0}D^{*0}}=&\int
 d\minv(D^{*0}K^{*0})\frac{1}{(2\pi)^3}\frac{|\vec{\tilde{p}}_{K^-}||\vec{\tilde{p}}\,^\prime_{K^{*0}}|}{4m_{B^-}^2}\sumpol|t_{B^-,K^-K^{*0}D^{*0}}|^2
 \label{eq_rate2}
\end{align}
with $\minv(D^{*0}K^{*0})$ being the invariant mass of the $D^{*0}K^{*0}$ pair, while
$\vec{\tilde{p}}_{K^-}$ and $\vec{\tilde{p}}\,^\prime_{K^{*0}}$  are the momenta of the
$K^-$ in the $B^-$ rest frame and that of the $K^{*0}$ in the $D^{*0}K^{*0}$
CM frame, given by
\begin{align}
 |\vec{\tilde{p}}_{K^-}|=&\frac{1}{2m_{B^-}}\lambda^{1/2}(m_{B^-}^2,m_{K^-}^2,\minv^2(D^{*0}K^{*0})),\\
 |\vec{\tilde{p}}\,^\prime_{K^{*0}}|=&\frac{1}{2\minv(D^{*0}K^{*0})}\lambda^{1/2}(\minv^2(D^{*0}K^{*0}),m_{K^{*0}}^2,m_{D^{*0}}^2) \ .
\end{align}
Using Eq.~(\ref{eq_sum_pol}) for the $D^{*0}$ and $K^{\ast 0}$ polarization sums of Eq.~(\ref{eq_rate2}), we find
\begin{align}
 \sumpol | t_{B^-,K^-K^{*0}D^{*0}}|^2 = C^{\prime\,2}\,
 \sumpol|\vec{\epsilon}_{K^{*0}}\,\vec{\epsilon}_{D^{*0}}|^2=3 C^{\prime\,2} \ .\label{t_pol_ksds}
\end{align}
However, once again, the momentum of the $K^{*0}$ is not small compared to the
$K^{*0}$ mass in the region of invariant $D^{*0}K^{*0}$ masses covered by the phase space of the process. One must then use the covariant form of
Eq.~(\ref{eq_gbmkmks0ds0}),
\begin{align}
 -{\rm i}t_{B^-,K^-K^{*0}D^{*0}}={\rm i}\,C^{\prime}\epsilon_{K^{*0}\mu}\epsilon^\mu_{D^{*0}},\label{eq_tbmdecds_re}
\end{align}
finding
\begin{align}
 \sumpol|t_{B^-,K^-K^{*0}D^{*0}}|^2=&\,C^{\prime\,2}\left(2+\frac{\left(\minv^2(D^{*0}K^{*0})-m_{D^{*0}}^2-m_{K^{*0}}^2\right)^2}{4m^2_{K^{*0}}m^2_{D^{*0}}}\right).
\end{align}
Then,  the constant $C^{\prime\,2}$ can be determined from
\begin{align}
 C^{\prime\,2}=&\frac{\Gamma_{B^-\rightarrow K^-K^{*0}D^{*0}}}{\displaystyle\int
 d\minv(D^{*0}K^{*0})\displaystyle\frac{1}{(2\pi)^3}\displaystyle\frac{|\vec{\tilde{p}}_{K^-}| |\vec{\tilde{p}}\,^\prime_{K^{*0}}|}{4m_{B^-}^2}\left(2+\frac{\left(\minv^2(D^{*0}K^{*0})-m_{D^{*0}}^2-m_{K^{*0}}^2\right)^2}{4m^2_{K^{*0}}m^2_{D^{*0}}}\right)} \ .
 \label{eq_cprimesq}
\end{align}

Similarly to the previous case, we rely on results from the chiral unitary approach \cite{daniaxial} to obtain the coupling constant of the $D_{s1}^+$ state to a $D^{*0}K^+$ pair.
This coupling is determined from the one of  $D_{s1}^+$ to $D^*K$ in isospin $I=0$ quoted in  \cite{daniaxial} ,
$g_{D_{s1},D^*K}=9.82$~GeV, multiplied by the appropriate CG coefficient. Hence,
$g_{D_{s1}^+,D^{*0}K^+}=-g_{D_{s1},D^*K}/\sqrt{2}$. The corresponding  $S$-wave amplitude
is given by
\begin{align}
 -{\rm i}\,t_{D_{s1}^+,D^{*0}K^+}=-{\rm i}\,g_{D^+_{s1},D^{*0}K^+}\vec{\epsilon}_{D_{s1}^+}\,\vec{\epsilon}_{D^{*0}},\label{eq_gds1ds0kp}
\end{align}
where $\vec{\epsilon}_{D_{s1}^+}$ and $\vec{\epsilon}_{D^{*0}}$ are the
polarization vectors of the $D_{s1}^+$ and $D^{*0}$, respectively.
In the region of the invariant masses where the singularity appears, the
momentum of the $D_{s1}^+$ is of the order of 10\% of its mass and, again, it is proper
to neglect the $\epsilon^0_{D_{s1}^+}$ component.

Using the expressions for the vertices given by Eqs.~(\ref{eq_gkskpi}), (\ref{eq_gbmkmks0ds0}), and (\ref{eq_gds1ds0kp}), the
amplitude of the triangle diagram of Fig.~\ref{fig_4} for the decay
$B^-\rightarrow K^-\pi^-D_{s1}^+$ in the $\pi^-D_{s1}^+$ CM frame is
given by
\begin{align}
 t_{B^-,K^-\pi^-D_{s1}^+}=&\,C^\prime\, g\, g_{D^+_{s1},D^{*0}K^+}\sum_{{\rm pol}(K^{*0},D^{*0})} 
 {\rm i}\int\frac{d^4q}{(2\pi)^4}\, (\vec{\epsilon}_{K^{*0}} \, \vec{\epsilon}_{D^{*0}})\,(\vec{\epsilon}_{K^{*0}} \, (2\vec{p}\,^\prime_{\pi^-}+\vec{q}\,))\,(\vec{\epsilon}_{D^{*0}}\, \vec{\epsilon}_{D_{s1}^+})\notag\\
 &\frac{1}{q^2-m_{D^{*0}}^2+{\rm i}\epsilon}\frac{1}{(P^\prime-q)^2-m_{K^{*0}}^2+{\rm i}\epsilon}\frac{1}{(P^\prime-q-p^\prime_{\pi^-})^2-m_{K^+}^2+{\rm i}\epsilon}\notag\\
 \equiv&\,C^\prime\, g\, g_{D^+_{s1},D^{*0}K^+}(\vec{p}\,^\prime_{\pi^-} \,\vec{\epsilon}_{D_{s1}^+})\,\tilde{t}_T'\ ,\label{eq_t_2}
\end{align}
where Eq.~(\ref{eq_sum_pol}) has been employed to sum over the polarizations of the $K^{*0}$ and the $D^{*0}$. The last identity of the previous equation defines the triangle singularity integral $\tilde{t}_T'$, which is formally the same as $\tilde{t}_T$ in Eq.~(\ref{eq_tT1})
with the replacement of $m_{D^0}$ by $m_{D^{*0}}$, 
$P^{0\prime}$ by $\minv(\pi^- D^+_{s1})$
and
\begin{align}
|\vec{p}\,^\prime_{\pi^-}|=\frac{1}{2\minv(\pi^-D^+_{s1})}\lambda^{1/2}(\minv^2(\pi^-D^+_{s1}),m_{\pi^-}^2,m_{D_{s1}^+}^2) \ .
\label{eq:pimom_2}
\end{align}
Following similar steps as in the previous case,
the mass distribution of the branching ratio for $B^-\rightarrow K^-\pi^-D_{s1}^+$ decay is given by
\begin{align}
 \frac{1}{\Gamma_B}\frac{d\Gamma_{B^-\rightarrow
 K^-\pi^-D_{s1}^+}}{d\minv(\pi^-
 D^+_{s1})}=&\frac{{\rm BR}(B^-\rightarrow
 K^-K^{*0}D^{*0})\cdot|\vec{p}_{K^-}||\vec{p}\,^\prime_{\pi^-}||\bar{t}^{\rm
 eff}_{B^-,K^-\pi^-D^+_{s1}}|^2}{\displaystyle\int
 d\minv(D^{*0}K^{*0})|\vec{\tilde{p}}_{K^-}||\vec{\tilde{p}}\,^\prime_{K^{*0}}|\left(2+\frac{\left(\minv^2(D^{*0}K^{*0})-m_{D^{*0}}^2-m_{K^{*0}}^2\right)^2}{4m^2_{K^{*0}}m^2_{D^{*0}}}\right)}
 \ ,\label{eq_mass_dist_2}
\end{align}
where $\vec{p}_{K^-}$ is
the $K^-$ momentum in the $B^-$ rest frame, given by
\begin{align}
 |\vec{p}_{K^-}|=&\frac{1}{2m_{B^-}}\lambda^{1/2}(m_{B^-}^2,m_{K^-}^2,\minv^2(\pi^-
 D^+_{s1})) \ ,
\end{align}
and we have used
\begin{align}
 \sumpol( \vec{p}\,^\prime_{\pi^-} \vec{\epsilon}_{D^+_{s1}})^2=|\vec{p}\,^\prime_{\pi^-}|^2 \ ,
\end{align}
which is readily obtained employing again Eq.~(\ref{eq_sum_pol}). In Eq.~(\ref{eq_mass_dist_2}) we have defined
$\bar{t}^{\rm eff}_{B^-,K^-\pi^-D_{s1}^+}=g\,g_{D_{s1}^+,D^{*0}K^+}\, |\vec{p}\,^\prime_{\pi^-}|\, \tilde{t}'_T$,
and the branching ratio ${\rm
BR}(B^-\rightarrow K^-K^{*0}D^{*0})=\Gamma_{B^-\rightarrow
K^-K^{*0}D^{*0}}/\Gamma_B$ is $(1.5\pm0.4)\times 10^{-3}$ \cite{pdg}. The integral of Eq.~(\ref{eq_mass_dist_2}) over
$\minv(\pi^-D^+_{s1})$ gives the branching ratio of the $B^-\rightarrow
K^-\pi^-D_{s1}^+$ process.

\section{Results}
\begin{figure}[t]
 \centering
 \includegraphics[width=12cm]{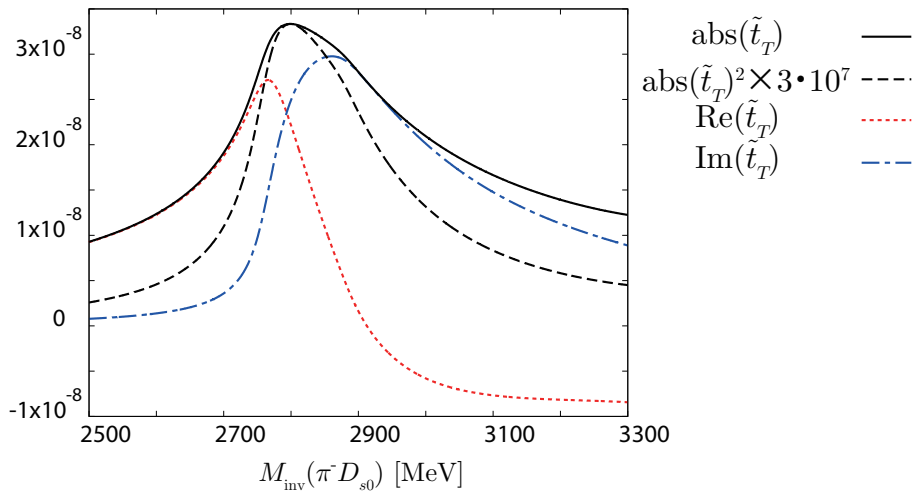}
 \caption{Triangle amplitude for the $B^-\rightarrow K^-\pi^-D_{s0}^+$
 process, $\mid\tilde{t}_T\mid$, $\mid\tilde{t}_T\mid^2$, Re$(\tilde{t}_T)$, and
 Im$(\tilde{t}_T)$ as functions of $\minv(\pi^-D_{s0}^+)$.
 The unit of $\tilde{t}_T$ is MeV$^{-2}$.}
 \label{fig_5}
\end{figure}

\begin{figure}[t]
 \centering
 \includegraphics[width=12cm]{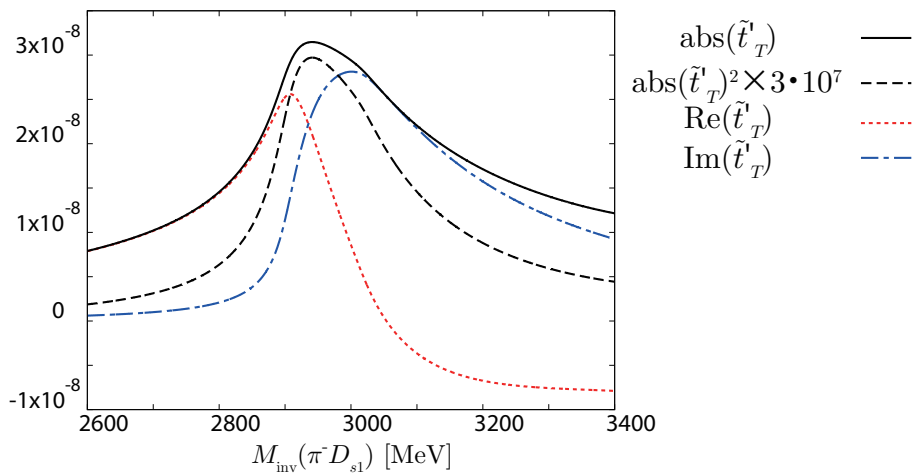}
 \caption{Triangle amplitude for the $B^-\rightarrow K^-\pi^-D_{s1}^+$
 process, $\mid\tilde{t}'_T\mid$, $\mid\tilde{t}'_T\mid^2$, Re$(\tilde{t}'_T)$, and
 Im$(\tilde{t}'_T)$ as functions of $\minv(\pi^-D_{s1}^+)$.
 $\tilde{t}'_T$ is given in the unit of MeV$^{-2}$.}
 \label{fig_6}
\end{figure}

In Fig.~\ref{fig_5}, we show the absolute value, square of the absolute value,
real part, and imaginary part of the triangle amplitude $\tilde{t}_T$ in
Eq.~(\ref{eq_tT1}) for the $B^-\rightarrow K^-\pi^-D_{s0}^+$ decay process, as
functions of the $\pi^-D_{s0}^+$ invariant mass.
The peak in $\mid\tilde{t}_T\mid^2$ is located at 2800~MeV and it has a width of about 200~MeV.
This width comes from the $K^*$ decay width and the fact that the condition
for the triangle singularity of Ref.~\cite{Bayar:2016ftu} cannot be
strictly fulfilled for the actual mass of the $D_{s0}^+(2317)$, because
this state is bound with respect to the $DK$ threshold at 2359~MeV and the $DK$ pair cannot be placed on-shell.
Yet, one can see that the peak is related to a ``nearly missed'' triangle
singularity by taking values of the $D_{s0}^+(2317)$ mass just above the
$DK$ threshold.
If we take a mass of the $D_{s0}^+(2317)$ close to the $DK$ threshold,
from $2360$~MeV to $2365$~MeV, the condition of Eq.~(18) of
Ref.~\cite{Bayar:2016ftu} gives the singularity between $2827$~MeV and
$2801$~MeV, close to the position of the peak in Fig.~\ref{fig_5}.
The results of Fig.~\ref{fig_6}, shown as
functions of $\minv(\pi^-D_{s1}^+)$, are the equivalent ones for the $\tilde{t}'_T$ amplitude in Eq.~(\ref{eq_t_2}), corresponding to the
$B^-\rightarrow K^-\pi^-D_{s1}^+$ decay process.
One can see a peak around 2950~MeV with a 150~MeV width.
The peak position is also similar to the value expected from the formula
in Ref.~\cite{Bayar:2016ftu}, which
ranges between $2971$~MeV and $2943$~MeV when we take a $D_{s1}^+$ mass
from $2502$~MeV to $2507$~MeV just above the $D^{*0}K^+$ threshold ($2501$~MeV).
Note that,
around the peaks, these amplitudes could be described by an expression of the type ${\rm i} BW(\minv)+{\rm i} B$, with
$BW(\minv)=1/(\minv^2-m_R^2+{\rm i} \minv\Gamma)$ and $B$ as a real background, therefore resembling the behavior expected for a hadron resonance.

\begin{figure}[t]
 \centering
 \includegraphics[width=17cm]{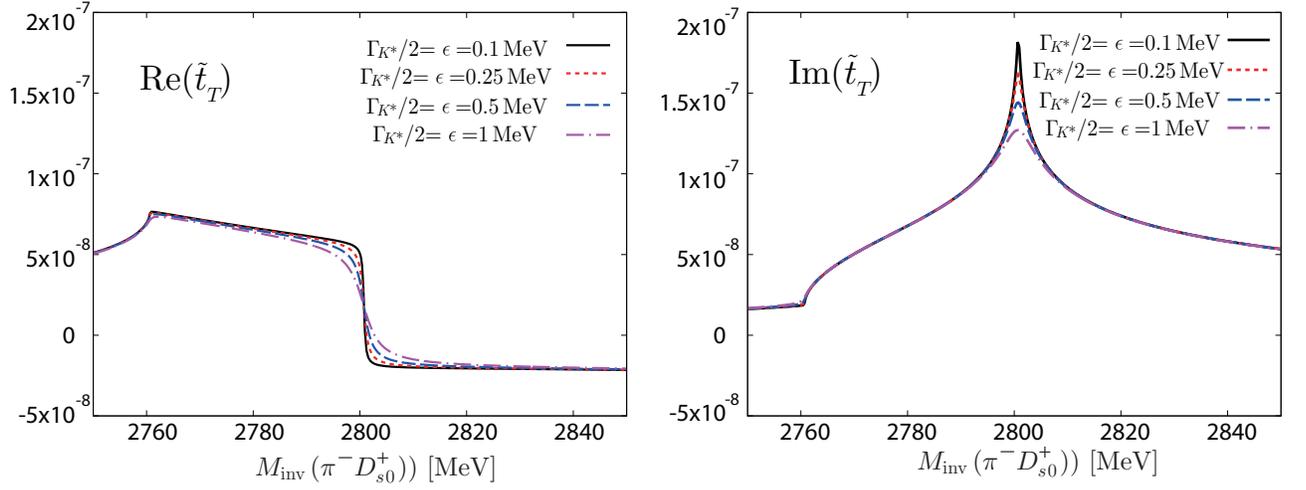}
 \caption{Real (left) and imaginary (right) parts of $\tilde{t}_T$ in Eq.~(\ref{eq_tT1}), with $m_K$ replaced by $m_K-60$~MeV and several
 values of  $\Gamma_{K^*}/2=\epsilon=1$, $0.5$, $0.25$ and $0.1$~MeV.}
 \label{fig_new}
\end{figure}

Before proceeding further, we would like to perform a few tests that allow us to interpret properly the peaks that appear both in the real and imaginary parts of $\tilde{t}_T$ and $\tilde{t}^\prime_T$
displayed in Figs.~\ref{fig_5} and \ref{fig_6}, respectively.
Let us discuss the case of Fig.~\ref{fig_5} since the one of Fig.~\ref{fig_6} would
be identical.
A triangle singularity appears in the diagram of Fig~\ref{fig_3}
when the $K^{*0}$, $D^0$, and $K^+$ particles are placed on-shell and
$\vec{p}\,'_\pi$ and $-\vec{q}$ are parallel ($K^{*0}$ and $\pi^-$ go in the
same direction), namely when
 $q_{\rm on}$ and $q_{a-}$ in
the nomenclature of Ref.~\cite{Bayar:2016ftu} are equal (see Eq.~(18) of this reference).
As mentioned before,
since the $D_{s0}^+(2317)$ is
bound with respect to the $D^0K^+$ component,  this meson pair
cannot be placed on-shell.
Hence, the condition $q_{\rm on}=q_{a-}$ cannot be strictly fulfilled.
However, since the binding is moderate, instead of the singularity that one would obtain (if 
$\Gamma_{K^*}$ was zero)  we find a finite enhancement.
As discussed before, the condition $q_{\rm on}=q_{a-}$ of
Ref.~\cite{Bayar:2016ftu} is fulfilled with masses of $D_{s0}^+(2317)$
slightly bigger than the $DK$ threshold. We have seen that the singularity moves towards higher energies as the 
$D_{s0}^+$ mass decreases and that at a mass of 2360~MeV, just 1 MeV above the $DK$ threshold, it appears at 2827~MeV.
It is then logical to associate the peak of Im$(\tilde{t}_T)$ at $2850$~MeV to
this ``nearly missed'' triangle singularity.
Note, however, that there is also a peak in the real part around $2758$~MeV that
instead corresponds to the threshold of the $K^{*0}D^0$ system. If the $K^{*0}$ had no width, this would have appeared as a cusp tied to having a $K^{*0}$ and a $D^0$ on-shell at threshold, while
the $K^+$ is off-shell. In order to corroborate this interpretation, we perform calculations where $\Gamma_{K^*}/2$ is artificially reduced while $\epsilon$ is taken finite but close to zero, which will help visualizing the building up of the singularity.
The results are shown in Fig.~\ref{fig_new}.
There we have also artificially reduced the mass of the kaon by $60$~MeV
so that the $D^0K^+$ mass is below the $D_{s0}^+(2317)$ mass by about
17~MeV, and hence the $K^{*0}$, $D^0$, and $K^+$ mesons can be placed simultaneously on-shell, producing a triangle singularity.
One can now see two structures in both $\re(\tilde{t}_T)$ and $\im(\tilde{t}_T)$: a cusp at the threshold of the 
$K^{*0}D^0$ state and the triangle singularity around 2800~MeV, seen as a narrow peak in $\im(\tilde{t}_T)$ and as a sharp downfall in  $\re(\tilde{t}_T)$.
In order to further stress this point, we take decreasing values of  
$\Gamma_{K^*}/2$ and $\epsilon$, namely
$1$, $0.5$, $0.25$ and $0.1$~MeV, and we then see that the cusp associated to the threshold in the real part
converges to a
finite value, while the peak of the singularity in the imaginary part keeps growing and becomes sharp as 
corresponds to a singularity.

\begin{figure}[ht]
 \centering
 \includegraphics[width=16cm]{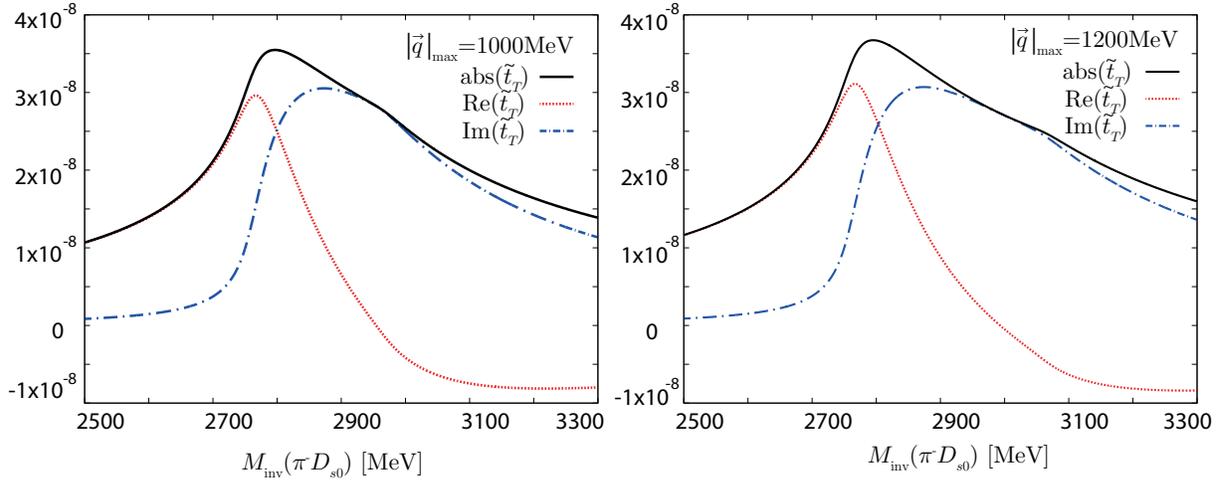}
 \caption{$\mid \tilde{t}_T\mid$, $\re(\tilde{t}_T)$ and $\im(\tilde{t}_T)$ with $|\vec{q}\,|_{\rm
 max}=1000$ and 1200 MeV.}
 \label{fig_new2}
\end{figure}

Coming back to Fig.~\ref{fig_5} for the realistic case, we should note
that as we increase $\minv(\pi^-D_{s0})$, so will the on-shell momentum
of $D^0K^{*0}$, and since we apply a momentum cutoff $|\vec{q}\,|_{\rm
max}$, this could impose constraints in the phase space and modify the shape of the calculated spectrum.
For this reason, we also investigate whether the falldown of $\im(\tilde{t}_T)$ is due to the
cutoff value employed or to the triangle singularity behavior.
We increase $|\vec{q}\,|_{\rm max}$ from the value $800$~MeV,
constrained by the unitary approach to obtain the $D_{s0}^+$ from the $DK$
component, to $1000$~MeV and 1200~MeV. The corresponding results are
shown in Fig.~\ref{fig_new2}.
We see that, as the cutoff value is increased,  the falldown of $\im(\tilde{t}_T)$ at higher
$\minv(\pi^-D_{s0})$ becomes softer, but in all cases the peak that we
have associated to the triangle singularity remains.

\begin{figure}[ht]
 \centering
 \includegraphics[width=9cm]{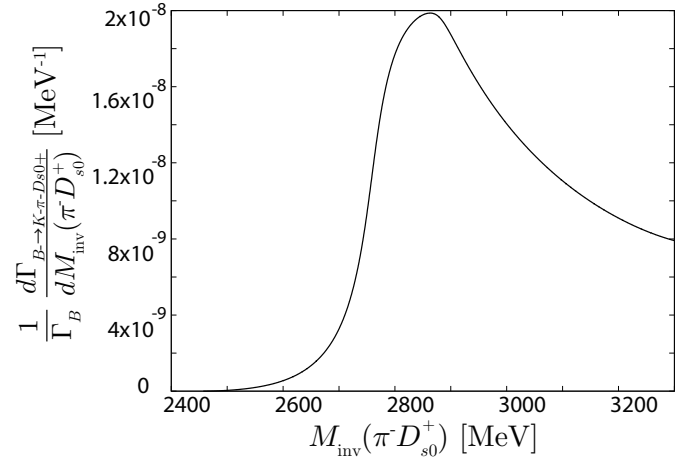}
 \caption{Mass distribution of the $B^-\rightarrow K^-\pi^-D_{s0}^+$
 decay as a function of $\minv(\pi^-D_{s0}^+)$.}
\label{fig_7}
\end{figure}

\begin{figure}[ht]
 \centering
 \includegraphics[width=9cm]{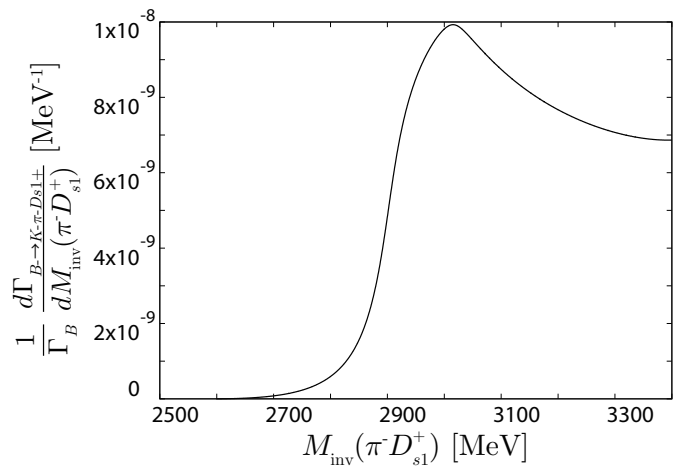}
 \caption{Mass distribution of the $B^-\rightarrow K^-\pi^-D_{s1}^+$
 decay as a function of $\minv(\pi^-D_{s1}^+)$.}
 \label{fig_8}
\end{figure}

In Fig.~\ref{fig_7}, we show the mass distribution of the
$B^-\rightarrow K^-\pi^-D_{s0}^+$ decay as a function of
$\minv(\pi^-D_{s0}^+)$ given in Eq.~(\ref{eq_mass_dist_1}).
A peak appears around 2850~MeV, reflecting the behavior observed in
Fig.~\ref{fig_5} for $\tilde{t}_T$ but with the additional weight of some kinematical factors. Analogously, 
Fig.~\ref{fig_8} shows the mass distribution of the
$B^-\rightarrow K^-\pi^-D_{s1}^+$ decay,  given in Eq.~(\ref{eq_mass_dist_2}), as a function of
$\minv(\pi^-D_{s1}^+)$. The peak of the mass distribution in this case is located at 3000~MeV, which lies somewhat above the peak of the corresponding $\tilde{t}'_T$ amplitude shown in  Fig.~\ref{fig_6}.
Upon inspecting the final results in Fig.~\ref{fig_7} for
$d\Gamma_{B^-\rightarrow K^-\pi^-D_{s0}^+}/d\minv(\pi^-D_{s0}^+)$, we note
that the peak around $2850$~MeV corresponds to where $|\tilde{t}_T|^2$ gets most of its
strength from $\im(\tilde{t}_T)$, the triangle singularity amplitude displayed in Fig.~\ref{fig_5},
and hence we can associate the structure seen in Fig.~\ref{fig_7} (and
Fig.~\ref{fig_8} for the $K^-\pi^-D_{s1}^+$ decay) mostly to the effect of the
triangle singularity.
 
We also integrate the mass distributions of Figs.~\ref{fig_7} and
\ref{fig_8} up to 400 MeV above the peak, and
we obtain a branching ratio of $7.8\times 10^{-6}$ for the
$B^-\rightarrow K^-\pi^-D_{s0}^+$ decay process and of  $4.2\times 10^{-6}$ for the $B^-\rightarrow K^-\pi^-D_{s1}^+$ one. These branching ratios
have not yet been reported in the PDG and, since they
are likely to be measured in a near future,
our prediction and the anticipation on establishing the nature of these unavoidable peaks, as coming from triangle singularities, is
most opportune.  Note, however, that the strength under the peak of the $B^-\rightarrow K^-\pi^-D_{s0}^+$ and $B^-\rightarrow K^-\pi^-D_{s1}^+$ distributions depends on the constants $C^2$ [Eq.~(\ref{eq_csq})] and $C^{\prime 2}$ [Eq.~(\ref{eq_cprimesq})] , respectively, which in turn depend on the particular shape adopted for the respective vertices, Eqs.~(\ref{eq_tbmdec_rel}) and (\ref{eq_tbmdecds_re}). Therefore, our final results for the branching ratios of the $B^-\rightarrow K^-\pi^-D_{s0}^+$ and $B^-\rightarrow K^-\pi^-D_{s1}^+$ processes under their respective peaks will be given in the next section, after implementing a more realistic form of the decay vertex, which accounts for the dominance of the $a_1(1260)$ resonance coupling strongly to a $K^- K^{*0}$ pair.


One may  wonder about
{possible sources} of background
{to the  $K^-\pi^-D_{s0}^+(2317)$
or $K^-\pi^-D_{s1}^+(2460)$ processes}.
In this sense,
{these reactions are special}.
{Indeed, one notes} that there is no resonance
with strangeness and charge $-2$ that could decay to $K^-\pi^-$.
On the other hand, one might also wonder what would happen if
{a}
$K^-DK'^*$
{intermediate state was produced,} with $K'^*$  being any resonance which can decay to $\pi K$, and
then we
{added the corresponding}
triangle diagram
{contribution of Fig.~\ref{fig_3}, upon} substituting $K^*$ by
$K'^*$.
{First of all, we note that there is no evidence in the PDG \cite{pdg} for the processes $B^-\rightarrow
K^-K'^*D$, with $K'^*$ being any of the states which decay
into $\pi K$.
Hence, an evaluation of the strength of the corresponding triangle amplitudes would not
be possible. In any case, this would not be necessary here since, by testing Eq.~(18) of Ref.~\cite{Bayar:2016ftu} with $m_{D_{s0}^+}$
slightly larger than the $DK$ pair mass ($3$~MeV above the threshold), the singularity appears at the
values shown in Table~\ref{table_new} for different $K'^*$ states up to 2000 MeV, and all of them are much higher  than the
value around $2815$ MeV that we have obtained with our mechanism.}
\begin{table}[t]
 \caption{The energy value $\sqrt{s_{TS}}$ at which a triangle
 singularity appears in the process of Fig.~\ref{fig_3}, for different
 $K'^*$ resonances instead of $K^*(892)$ .
 The values of $\sqrt{s_{TS}}$ are evaluated taking
 $m_{D_{s0}^+}=2362$~MeV.}
 \label{table_new}
 \begin{tabular}[t]{c|c|c|c|c|c}
  &$K^*(1410)(1^-)$ &$K^*_0(1430)(0^+)$ &$K^*(1680)(1^-)$
	      &$K(1780)(3^-)$ &$K^*_0(1950)(0^+)$ \\\hline
  $\sqrt{s_{TS}}$ [MeV] &$3617$ &$3650$ &$4085$ &$4264$ &$4575$ \\
 \end{tabular}
\end{table}

We would also like to estimate possible sources of errors.
In the first place, for the evaluation of the triangle diagram we made the approximation
of neglecting the zeroth component of the $K^*$ polarization vector
$\epsilon^\mu(K^*)$ in Eq.~(\ref{eq_gkskpi}).
This was justified because the $K^*$ momentum for the peak of the
singularity is $262$~MeV$/c$, which is smaller than the $K^*$ mass.
For parity reasons, the omission of the $\epsilon^0$ component goes
in corrections
{of the type}
$(p_{K^*}/m_{K^*})^2$, which
{are} of  the order of $8\%$.
Actually, there are factors that make this error even smaller and{, as shown in the appendix, the final corrections turn to be of the order of $1.5\%$.}
Another source of uncertainty in our results is the coupling of the
$D_{s0}^+(2317)$ to $DK$ ($D_{s1}^+(2460)$ to $D^*K$).
From the analyses of Refs.~\cite{Gamermann:2006nm,sasa}, we estimate this uncertainty below
$10\%$, which would induce uncertainties of $20\%$ in our results.
However, the biggest uncertainty comes from the experimental error in the
$B^-\rightarrow K^-K^{*0}D^{*0}$ branching ratio $(1.5\pm 0.4)\times
10^{-3}$, which is $27\%$.
Summing all the uncertainties in quadrature we get $34\%$.
It is thus realistic to attribute some $30-40\%$ uncertainties to the
absolute values of the obtained rates.
Yet the shape of the distribution as a function of energy, which is the
distinctive feature of the triangle singularity, should be much more
accurate.


\section{Further considerations}
\label{sec:reson}
The position of a triangle singularity is of kinematical origin, tied to
the possibility to have $K^{*0}D^0K^+$ $(K^{*0}D^{*0}K^+)$ on-shell with collinear
momenta, and hence not strongly dependent on the $B^-\rightarrow
K^-D^0K^{*0}$ $(K^-D^{*0}K^{*0})$ vertex.
We have given a suitable structure for these vertices based on angular
momentum conservation by means of Eqs.~(\ref{eq_tbmdec_rel}) and
(\ref{eq_tbmdecds_re}).
In practice, it could be more complex.
In this section, we examine what happens if we consider that this
primary decay is dominated by the $a_1(1260)$ resonance coupled to
$K^-K^{*0}$, as implied by the $K^-K^{*0}$ invariant mass distribution of
Ref.~\cite{belle} (see Fig.~3 (a) of this reference).
Actually, in that work the invariant mass distribution is given for the combined
$B^-\rightarrow K^-D^0K^{*0}$ and $K^-D^{*0}K^{*0}$ decays.
We discuss here what happens if we introduce also a dominance of the
$a_1(1260)$ resonance for the $K^-K^{*0}$ system.
We make it by multiplying the amplitudes that we have in
Eqs.~(\ref{eq_tbmdec_rel}) and (\ref{eq_tbmdecds_re}) by an extra factor
accounting for the possible $a_1(1260)$ resonant shape in the
$K^-K^{*0}$ invariant mass distribution,
\begin{align}
 B_W(\minv(K^-K^{*0}))=\frac{m_{a_1}\Gamma_{a_1}}{\minv^2(K^-K^{*0})-m_{a_1}^2+im_{a_1}\Gamma_{a_1}},\label{eq_new22}
\end{align}
and, as in Ref.~\cite{belle}, we take $m_{a_1}=1230$~MeV and
$\Gamma_{a_1}=460$~MeV.
This requires now some changes in the formalism.
Equation~(\ref{eq:gamma1}) is now changed to
\begin{align}
 \Gamma_{B^-\rightarrow K^-K^{*0}D^0}=&\int
 d\minv(K^-K^{*0})\frac{1}{(2\pi)^3}\frac{|\vec{\tilde{p}}_{D^0}||\vec{\tilde{p}}\,''_{K^-}|}{4m_{B^-}^2}\sum_{\rm
 pol}|t_{B^-,K^-K^{*0}D^0}|^2|B_W(\minv(K^-K^{*0}))|^2,
\end{align}
where the integration runs over the variable on which both the amplitude and $B_W$ depend, and
 $\vec{\tilde{p}}_{D^0}$ and $\vec{\tilde{p}}\,''_{K^-}$ are the
momenta of $D^0$ in the $B^-$ rest frame and $K^-$ in the $K^-K^{*0}$
CM frame given by 
\begin{align}
 |\vec{\tilde{p}}_{D^0}|=&\frac{\lambda^{1/2}(m^2_{B^-},m^2_{D^0},\minv^2(K^-K^{*0}))}{2m_{B^-}},\\
 |\vec{\tilde{p}}\,''_{K^-}|=&\frac{\lambda^{1/2}(\minv^2(K^-K^{*0}),m_{K^-}^2,m_{K^{*0}}^2)}{2\minv(K^-K^{*0})},\label{eq_new21}
\end{align}
respectively.
The polarization sum of the square of the matrix element is given by
\begin{align}
 \sum_{\rm pol}|t_{B^-,K^-K^{*0}D^0}|^2=&C''^2\left(-m_{K^-}^2+\frac{(p_{K^-}\cdot p_{K^{*0}})^2}{m_{K^{*0}}^2}\right)
\end{align}
with $p_{K^-}\cdot
p_{K^{*0}}=[\minv^2(K^-K^{*0})-m_{K^-}^2-m_{K^{*0}}^2]/2$.
Similarly, Eq.~(\ref{eq_rate2}) is changed to
\begin{align}
 \Gamma_{B^-\rightarrow K^-K^{*0}D^{*0}}=&\int
 d\minv(K^-K^{*0})\frac{1}{(2\pi)^3}\frac{|\vec{\tilde{p}}_{D^{*0}}||\vec{\tilde{p}}\,''_{K^-}|}{4m_{B^-}^2}\sum_{\rm
 pol}|t_{B^-,K^-K^{*0}D^{*0}}|^2|B_W(\minv(K^-K^{*0}))|^2
\end{align}
with the $D^{*0}$ momentum in the $B^-$ rest frame,
\begin{align}
 |\vec{\tilde{p}}_{D^{*0}}|=\frac{\lambda^{1/2}(m_{B^-}^2,m_{D^{*0}}^2,\minv^2(K^-K^{*0}))}{2m_{B^-}}.
\end{align}
The square of the $B^-\rightarrow K^-K^{*0}D^{*0}$ matrix element with
the polarization sum is given by
\begin{align}
 \sum_{\rm pol}|t_{B^-,K^-K^{*0}D^{*0}}|^2=&C'''^2\left(2+\frac{(\tilde{p}\,''^0_{D^{*0}}\tilde{p}\,''^0_{K^{*0}})^2+\frac{1}{3}(|\vec{\tilde{p}}\,''_{D^{*0}}||\vec{\tilde{p}}\,''_{K^{*0}}|)^2}{m_{D^{*0}}^2m_{K^{*0}}^2}\right)
\end{align}
where $\vec{\tilde{p}}\,''_{D^{*0}}$
is the momentum of the $D^{*0}$ in the $K^-K^{*0}$ CM frame,
\begin{align}
 |\vec{\tilde{p}}\,''_{D^{*0}}|=\frac{\lambda^{1/2}(m_{B^-}^2,m_{D^{*0}}^2,\minv^2(K^- K^{*0}))}{2\minv(K^- K^{*0})};\
 \tilde{p}\,''^0_{D^{*0}}=\sqrt{m_{D^{*0}}^2+|\vec{\tilde{p}}\,''_{D^{*0}}|^2},
\end{align}
with $|\vec{\tilde{p}}\,''_{K^{*0}}|$ given by Eq.~(\ref{eq_new21}) and
$\tilde{p}\,''^0_{K^{*0}}=\sqrt{m_{K^{*0}}^2+|\vec{\tilde{p}}\,''_{K^{*0}}|^2}$.
In the triangle diagram, which we evaluate in the CM frame of
$\pi^-D_{s0}^+$ $(\pi^-D_{s1}^+)$, we have to multiply the integrand of
$t_T$ by $B_W(\minv(K^-K^{*0}))$ which is given by Eq.~(\ref{eq_new22}) with
\begin{align}
 \minv^2(K^- K^{*0})=&(p_{K^-}'^0+P'^0-q^0)^2-|\vec{p}\,'_{K^-}-\vec{q}\,|^2
\end{align}
In the loop integration of Eq.~(\ref{eq_tT3}), $q^0$ becomes
$\omega_D(\vec{q}\,)$ $(\omega_{D^*}(\vec{q}\,))$ by taking the
relevant positive energy parts of the intermediate propagators.
Then, we have for the $K^-\pi^-D_{s0}^+$ production (the same for the
$K^-\pi^-D_{s1}^+$ production by changing $D_{s0}^+\rightarrow D_{s1}^+$ and
$\omega_D(\vec{q}\,)\rightarrow \omega_{D^*}(\vec{q}\,)$)
\begin{align}
 \minv^2(K^-K^{*0})=&(p_{K^-}'^0+\minv(\pi^-D_{s0}^+)-\omega_D(\vec{q}\,))^2-\left(|\vec{p}\,'_{K^-}|^2+|\vec{q}\,|^2-2\vec{p}\,'_{K^-}\cdot\vec{q}\,\right),
\end{align}
where $\vec{p}\,'_{K^-}$ is the momentum of the $K^-$ in the CM
frame of $\pi^-D_{s0}^+$.
We can see now that we have a dependence on a new angle, the one between
$\vec{p}\,'_{K^-}$ and $\vec{q}$ which we define as $\theta_{K^-,q}$.
Rather than redoing the formulation including three new angular integrals,
we make an approximation of omitting the linear term in $|\vec{q}\,|$
because $|\vec{q}\,|$ is much smaller than $|\vec{p}\,'_{K^-}|$.
Indeed, for the relevant $\pi^-D_{s0}^+$ invariant mass where the
singularity appears we have $|\vec{p}_{K^-}|=3386$~MeV and
$|\vec{q}\,|=335$~MeV.
This, plus the $\cos\theta_{K^-,q}$ angle dependence, justifies such an
approximation.
It essentially corresponds to taking $\cos\theta_{K^-,q}=0$, but we
check also the results
for other angles, and make an average and evaluate the errors.

\begin{figure}[t]
 \centering
 \includegraphics[width=7cm]{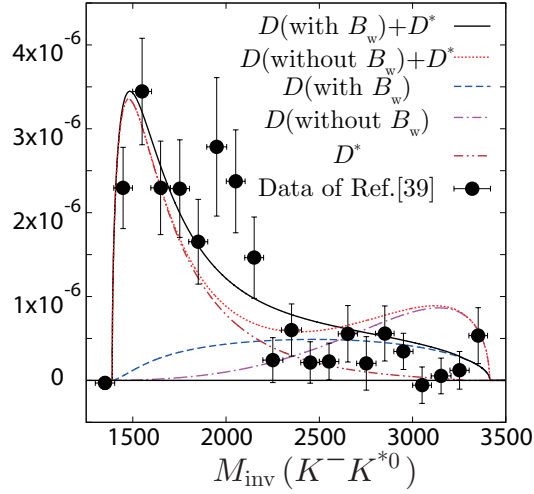}
 \caption{The mass distributions $d\Gamma/d\minv(K^-K^{*0})$ for the $B^-\rightarrow K^-K^{*0}D^0$
 and $K^-K^{*0}D^{*0}$ processes, and their sum as functions of
 $\minv(K^-K^{*0})$.
 The normalization of the data $dN/d\minv(K^-K^{*0})$ from
 Ref.~\cite{belle} has been fixed at
 the peak to agree with the calculated $d\Gamma/d\minv(K^-K^{*0})$.}
 \label{fig_anew}
\end{figure}

In Fig.~\ref{fig_anew}, we show the mass distributions
for the $B^-\rightarrow K^-K^{*0}D^0$ and $B^-\rightarrow
K^-K^{*0}D^{*0}$ reactions as functions of $\minv(K^-K^{*0})$.
Since in the experiment of Ref.~\cite{belle} the $K^-K^{*0}$ mass
distribution for the sum of the two decays is shown, we also sum the two
distributions, but weighted by their respective branching ratios,
$7.5\times 10^{-4}$ for $B^-\rightarrow K^-K^{*0}D^0$ and $1.5\times
10^{-3}$ for $B^-\rightarrow K^-K^{*0}D^{*0}$.
It is curious that the second reaction shows a neat signal for the tail
of the $a_1(1260)$, but this is not the case for the $B^-\rightarrow
K^-K^{*0}D^0$.
This is due to the $\vec{p}_{K^-}$ factor in Eq.~(\ref{eq_gbmkmks0d0}) that
peaks where the $K^-K^{*0}$ invariant mass is large.
In the absence of the $B_W$ factor, the mass distribution has a dip in
the low $K^-K^{*0}$ invariant mass region.
It is noteworthy that the weighted sum of the two distributions
agrees well with the experimental distribution of Ref.~\cite{belle}.
It would be very interesting to disentangle these two distributions to
see if what we obtain agrees with the data, or one should rather
construct more elaborated vertices.
Actually, with the $\vec{p}_{K^-}$ momentum dependence in
Eq.~(\ref{eq_gbmkmks0d0}) which we have justified from the microscopic
picture for the decay, it is not easy to generate the $K^*\bar{K}$
$s$-wave $a_1(1260)$, so we also show the results for the
$B^-\rightarrow K^-K^{*0}D^0$ decay in the absence of the resonance.
We see that the sum of the two distributions is not much different and
still compatible with the experiment except at very large invariant mass.

\begin{figure}[t]
 \centering
 \includegraphics[width=10cm]{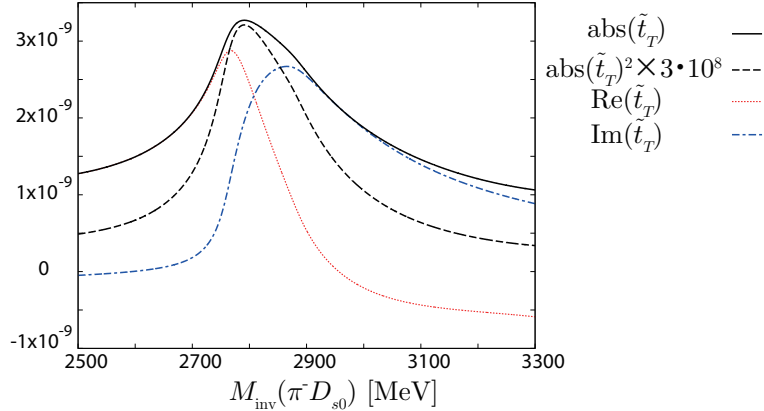}
 \caption{The triangle amplitude for the $B^-\rightarrow
 K^-\pi^-D_{s0}^+$ process including $B_W(\minv(K^-K^{*0}))$
 of Eq.~(\ref{eq_new22}).}
 \label{fig_newnew1}
\end{figure}

\begin{figure}[t]
 \centering
 \includegraphics[width=10cm]{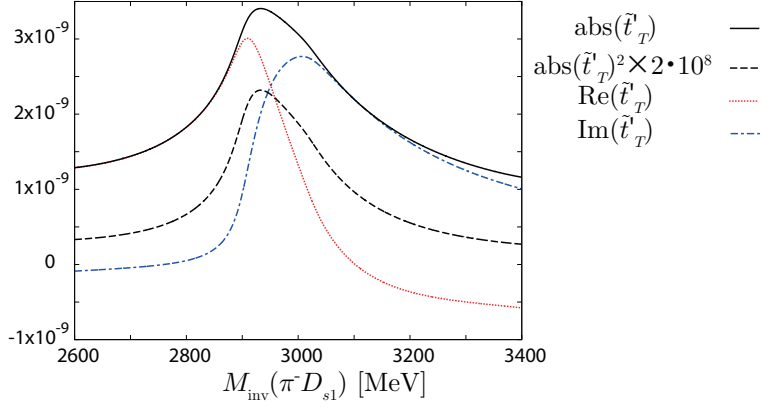}
 \caption{The triangle amplitude for the $B^-\rightarrow
 K^-\pi^-D_{s1}^+$ process including
 $B_W(\minv(K^-K^{*0}))$ of Eq.~(\ref{eq_new22}).}
 \label{fig_newnew2}
\end{figure}

In Fig.~\ref{fig_newnew1}, we show now the triangle amplitude for the
$B^-\rightarrow K^-\pi^-D_{s0}^+$ process corresponding to
Fig.~\ref{fig_5}.
The units are arbitrary as in Fig.~\ref{fig_5}.
We can see that there are some minor modifications in the shape, but the
structure of the triangle singularity remains.
In Fig.~\ref{fig_newnew2}, we show the triangle amplitude for the
$B^-\rightarrow K^-\pi^-D_{s1}^+$ corresponding to Fig.~\ref{fig_6}.
Once again, we can see the peak structure from the triangle
singularity with some minor modifications from $B_W(K^-K^{*})$ compared
with Fig.~\ref{fig_6}.

\begin{figure}[t]
 \centering
 \includegraphics[width=10cm]{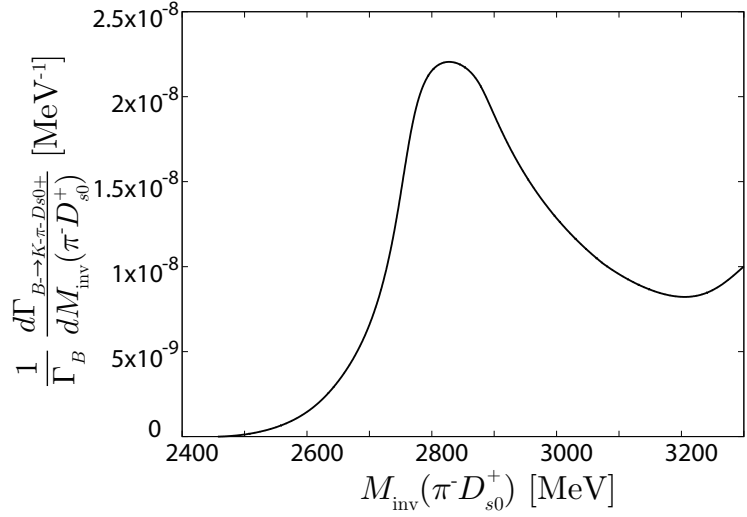}
 \caption{The mass distribution $d\Gamma_{B^-\rightarrow
 K^-\pi^-D_{s0}^+}/d\minv(\pi^-D_{s0}^+)$ as a function of
 $\minv(\pi^-D_{s0}^+)$ including $B_W(\minv(K^-K^{*0}))$ in
 Eq.~(\ref{eq_new22}).}
 \label{fig_newnew3}
\end{figure}

\begin{figure}[t]
 \centering
 \includegraphics[width=10cm]{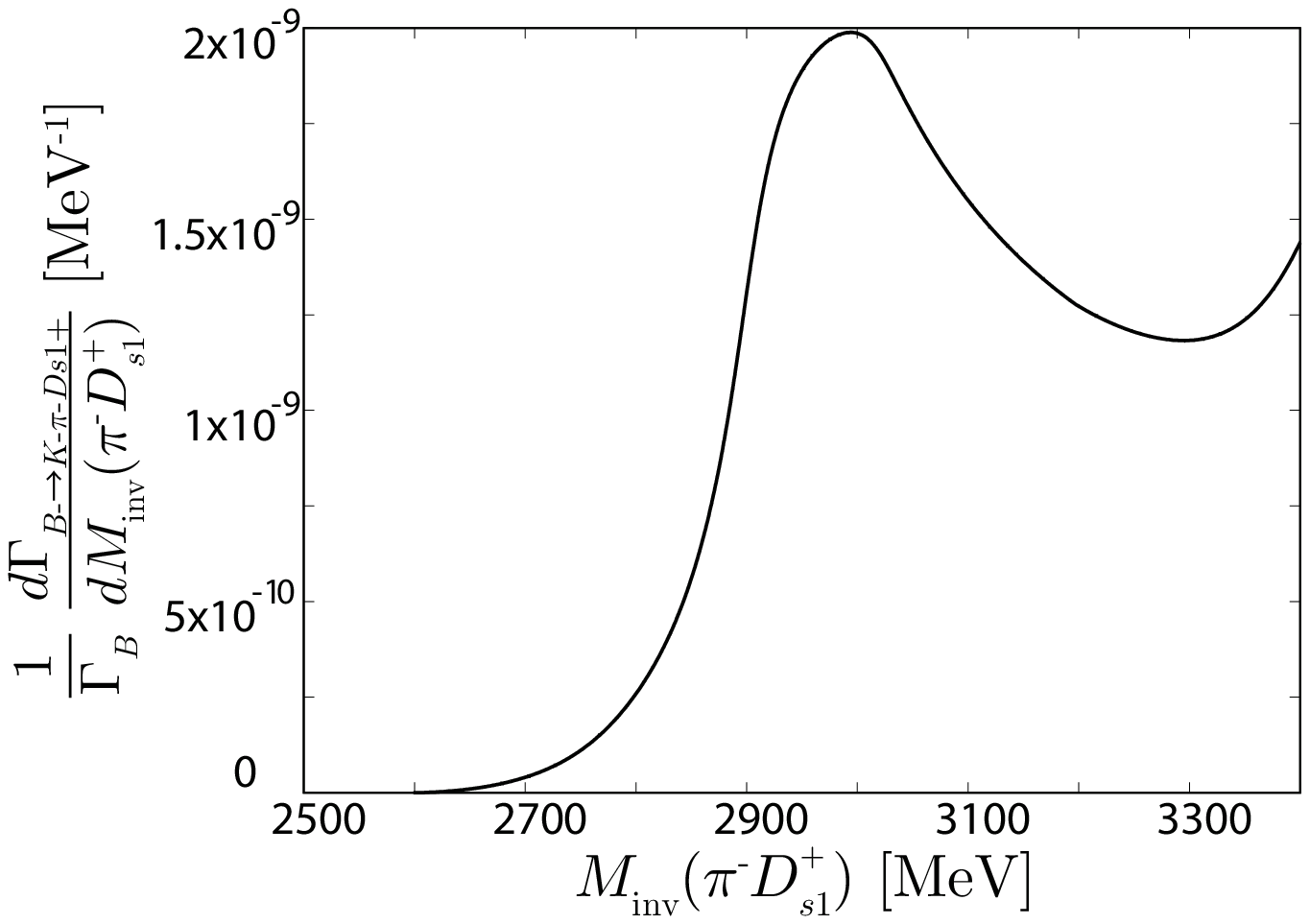}
 \caption{The mass distribution $d\Gamma_{B^-\rightarrow
 K^-\pi^-D_{s1}^+}/d\minv(\pi^-D_{s1}^+)$ as a function of
 $\minv(\pi^-D_{s1}^+)$ including $B_W(\minv(K^-K^{*0}))$ in
 Eq.~(\ref{eq_new22}).}
 \label{fig_newnew4}
\end{figure}

Finally, in Figs.~\ref{fig_newnew3} and \ref{fig_newnew4}
we show the mass distributions for the $B^-\rightarrow K^-\pi^-D_{s0}^+$
and $B^-\rightarrow K^-\pi^-D_{s1}^+$ processes including the $B_W(\minv(K^-K^{*0}))$
factor corresponding to Figs.~\ref{fig_7} and \ref{fig_8}, respectively.
We show the results for the average of $\cos\theta_{K^-,q}$.
Within the normalization obtained using the vertices of
Eq.~(\ref{eq_tbmdec_rel}) and (\ref{eq_tbmdecds_re}), the normalization
in the figures is absolute.
In Fig.~\ref{fig_newnew3}, we find a small variation in the size, but
the shape is similar to that obtained before.
In Fig.~\ref{fig_newnew4}, we find a similar behavior, but now the size
has been reduced by about a factor of five.
This reduces the integrated branching ratios to about $8\times10^{-6}$
and $1\times 10^{-6}$
with errors of about 60\%.
Yet, the message is clear.
The triangle singularities are very solid, and the absolute rates are
still within the measurable range. 

It is also important to stress that the strength of the
{$B^-\rightarrow K^-\pi^-D_{s0}^+$} reaction is
tied to the coupling of  the $D_{s0}^+(2317)$ to $DK$,
one important output of the picture in which the $D_{s0}^+(2317)$
would be basically a molecular state of $DK$.
{The same is applied to the reaction $B^-\rightarrow K^-\pi^-D_{s1}^+$
and the interpretation of the $D_{s1}^+(2460)$ as a $D^* K$ molecule.}
The measurement of
{these reactions} and comparison of
{their} strength to the
predictions done here would come to further support
{the molecular} hypothesis
{for these states}.

\section{Conclusion}
We have performed a study of the $B^-\rightarrow K^-\pi^-D_{s0}^+(2317)$ and
$B^-\rightarrow K^-\pi^-D_{s1}^+(2460)$ reactions and shown that they
develop a triangle singularity for an invariant mass of $2850$ MeV in
$\pi^-D_{s0}^+(2317)$ and $3000$ MeV in $\pi^-D_{s1}^+(2460)$,
respectively.
This triangle singularity shows up as a peak in the invariant mass
distribution of these pairs with an apparent width of about $200$ MeV.
The integrated strength in a region of about $400$ MeV around the peaks
gives branching rations {of} about $8\times10^{-6}$ and $1\times
10^{-6}$, respectively, which are within present measurable
range \cite{pdg}.

The singularity in the $B^-\rightarrow K^-\pi^-D_{s0}^+(2317)$ reaction
{is initiated by the transition} $B^-\rightarrow K^-K^{*0}D^0$, followed by the
$K^{*0}\rightarrow \pi^-K^+$ decay and the fusion of $K^+D^0$ to give
the $D_{s0}^+(2317)$ state.
The choice of the $D_{s0}^+(2317)$ in the final state is motivated by
the large coupling of $DK$ to this resonance, which in several
theoretical works, as well as from lattice QCD simulations, qualifies
mostly as a $DK$ molecule.
The second reaction, $B^-\rightarrow K^-\pi^-D_{s1}^+(2460)$, proceeds
in an analogous way, first the $B^-\rightarrow K^-K^{*0}D^{*0}$ {transition}
occurs, the $K^{*0}$ decays into $\pi^-K^+$ and the $K^+D^{*0}$ fuse to
{produce} the $D_{s1}^+(2460)$, which also, according to theoretical
calculations and lattice QCD simulations, corresponds to a molecular
state mostly composed of $D^*K$.

The predictions {found} here constitute a clear case of a peak produced by
a triangle singularity, which could be misidentified with a resonance
when the experiment is done.
{This} work has then the value of targeting a suitable reaction to identify
a triangle singularity, and then {having} the results and the study ready to
correctly interpret the peaks when they are observed.
We also stressed that the observation of those reactions would provide
further support for the molecular picture of the $D_{s0}^+(2317)$ and
$D_{s1}^+(2460)$ states.
With the steady advances in some of the experimental facilities,
particularly, with the LHCb and Belle experiments, we hope that these
measurements can be done in the near future, helping us get a better
understating of hadronic physics.

\section*{Acknowledgements}
We would like to thank De-Liang Yao for technical support.
This work is partly supported by the Spanish Ministerio de Economia y
Competitividad (MINECO) under the project SEV-2014-0398 of IFIC (Unidad
de Excelencia Severo Ochoa), the project MDM-2014-0369 of ICCUB (Unidad
de Excelencia María de Maeztu), the Spanish Excellence Network on
Hadronic Physics FIS2014- 57026-REDT, and, with additional European
FEDER funds, under the contracts FIS2011-28853-C02-01, FIS2011-24154 and
FIS2014-54762-P. Support has also been received from the Generalitat
Valenciana in the program Prometeo II-2014/068 and from the Generalitat
de Catalunya under contract 2014SGR-401.

\appendix*
\section{The approximation of neglecting the $\epsilon^0$ component in
 the $K^*$ polarization}
\label{sec_app}
{We take Eq.~(\ref{eq_gkskpi}) and we write the covariant form of it
\begin{align}
  -it'_{K^{*0},K^+\pi^-}=&+ig\epsilon_{K^{*0}\mu}(p_\pi-p_K)^\mu \  ,
\end{align}
then we calculate
\begin{align}
 \sumpol |t'|^2=&g^2\left(-g_{\mu\nu}+\frac{P^*_\mu
 P^*_\nu}{m_{K^*}^2}\right)(p_\pi -p_K)^\mu(p_\pi
 -p_K)^\nu=g^2(\vec{p}_\pi -\vec{p}_K)^2
\end{align}
where $P^*_\mu$, $p_{\pi\mu}$, and $p_{K\mu}$ are the momenta of $K^*$,
$\pi$, and $K$ in the rest frame of the $K^*$.
Now, if we take Eq.~(\ref{eq_gkskpi}),
\begin{align}
 -it_{K^{*0},K^+\pi^-}=&-ig\epsilon_{K^{*0}j}(p'_\pi -p'_K)_j,
\end{align}
we find
\begin{align}
 \sumpol |t|^2=g^2\delta_{ij}(p'_\pi -p'_K)_i(p'_\pi -p'_K)_j=g^2(\vec{p}\,'_\pi -\vec{p}\,'_K)^2
\end{align}
with $\vec{p}\,'_\pi$, $\vec{p}\,'_K$ evaluated in the $K^* D$ rest frame,
where $K^{*0}$ has $262$~MeV$/c$.
In order to calculate $\vec{p}\,'_\pi$ and $\vec{p}\,'_K$ we make a boost from
the rest frame of $K^{*0}$ to the one where it has a momentum
$|\vec{p}\,^*|=262$~MeV$/c$.
This boost is readily done and we get the result
\begin{align}
 \vec{p}\,'_\pi=&\left[\left(\frac{E^*}{m^*}-1\right)\frac{\vec{p}_\pi\cdot\vec{p}\,^*}{|\vec{p}\,^{*}|^2}+\frac{E_\pi}{m^*}\right]\vec{p}\,^*+\vec{p}_\pi
\end{align}
and similar for the $K$, where $m^*=m_{K^*}$,
$E^*=\sqrt{m^{*2}+|\vec{p}\,^*|^2}$, and $\vec{p}_\pi$, $E_\pi$ are the momentum of the
pion and its energy in the $K^*$ rest frame.
Upon integrating $\sum|t|^2$ over the solid angle, $\int d\Omega$, to get
the decay width of the $K^*$, we readily obtain
\begin{align}
 \frac{\int d\Omega \left(|t|^2-|t'|^2\right)}{\int
 d\Omega|t'|^2}=\frac{(E_\pi
 -E_K)^2}{4|\vec{p}_\pi|^2}\frac{|\vec{p}\,^{*}|^2}{m^{*2}}+O\left(\frac{|\vec{p}\,^{*}|^4}{m^{*4}}\right),
\end{align}
which gives an effect of the order of $1.5\%$.}

\end{document}